\def\lam{$\lambda$}
\def\degree{\hbox{$^\circ$}}
\def\arcmin{\hbox{$^\prime$}}
\def\arcsec{\hbox{$^{\prime\prime}$}}
\def\hour{\hbox{$^{\rm h}$}}
\def\min{\hbox{$^{\rm m}$}}

\def\fs{\hbox{$.\!\!^{\rm s}$}}

\def\farcs{\hbox{$.\!\!^{\prime\prime}$}}  
\def\lsim{\mathrel{\hbox{\rlap{\lower.55ex \hbox {$\sim$}}\kern-.0em
\raise.4ex \hbox{$<$}}}} 
\def\gsim{\mathrel{\hbox{\rlap{\lower.55ex \hbox {$\sim$}}\kern-.0em
\raise.4ex \hbox{$>$}}}} 
\def\lya{Ly$\alpha$}
\def\lyb{Ly$\beta$}
\def\grb{GRB\,030323}
\def\subsun{\mbox{$_{\odot}$}}
\def\ion#1#2{#1$\;${\small\rm\@Roman{#2}}\relax}

\documentclass{aa}
\usepackage{natbib,psfig,graphicx,rotating}

\bibpunct{(}{)}{;}{a}{}{,}

\begin{document}
\thispagestyle{empty}
\title{The host of \grb\ at $z$=3.372: a very high column density DLA
  system with a low metallicity\thanks{Based on observations collected
    at the European Southern Observatory, Chile, by GRACE (Gamma-Ray
    Burst Afterglow Collaboration at ESO), under program ID
    70.D-0523(B), and with NASA's Hubble Space Telescope.}}

\author{P.~M. Vreeswijk\inst{1}
\and
S.~L. Ellison\inst{1,2}
\and
C. Ledoux\inst{1}
\and
R.~A.~M.~J. Wijers\inst{3}
\and
J.~P.~U. Fynbo\inst{4,5}
\and
P. M{\o}ller\inst{6}
\and
A. Henden\inst{7}
\and
J. Hjorth\inst{5}
\and
G. Masi\inst{8}
\and
E. Rol\inst{3}
\and 
B.~L. Jensen\inst{5}
\and
N. Tanvir\inst{9}
\and
A. Levan\inst{10}
\and
J.~M. Castro Cer\'on\inst{11}
\and
J. Gorosabel\inst{12,11}
\and
A.~J. Castro-Tirado\inst{12}
\and
A.~S. Fruchter\inst{11}
\and
C. Kouveliotou\inst{13}
\and
I. Burud\inst{11}
\and
J. Rhoads\inst{11}
\and
N. Masetti\inst{14} 
\and
E. Palazzi\inst{14}
\and
E. Pian\inst{14,15}
\and
H. Pedersen\inst{5}
\and
L. Kaper\inst{3}
\and
A. Gilmore\inst{16}
\and
P. Kilmartin\inst{16}
\and
J.~V. Buckle\inst{17}
\and
M.~S. Seigar\inst{17}
\and
D.~H. Hartmann\inst{18}
\and
K. Lindsay\inst{18}
\and
E.~P.~J. van den Heuvel\inst{3}
}

\offprints{pvreeswi@eso.org}
   
\institute{European Southern Observatory, Alonso de C\'ordova 3107, 
Casilla 19001, Santiago 19, Chile
\and
P. Universidad Cat\'olica de Chile, Casilla 306, Santiago 22, Chile
\and
Astronomical Institute `Anton Pannekoek', University of Amsterdam \&
Center for High Energy Astrophysics, Kruislaan 403, 1098 SJ Amsterdam,
The Netherlands
\and
Department of Physics and Astronomy, {\AA}rhus
University, Ny Munkegade, DK-8000 {\AA}rhus C, Denmark
\and
Niels Bohr Institute, Astronomical Observatory, Copenhagen University,
Juliane Mariesvej 30, DK-2100 K{\o}benhavn {\O}, Denmark
\and
European Southern Observatory,
Karl-Schwarzschild-Strasse 2, D-85748, Garching bei M\"unchen, Germany
\and
Universities Space Research Association / U. S. Naval Observatory
Flagstaff Station, P.\,O. Box 1149, Flagstaff AZ 86002-1149, USA
\and
Physics Department, University of Rome "Tor Vergata", Via della
Ricerca Scientifica 1, 00133 Rome, Italy
\and
Department of Physical Sciences, University of Hertfordshire,
College Lane, Hatfield, Herts AL10 9AB, UK
\and
Department of Physics and Astronomy, University of Leicester, University
Road, Leicester, LE1 7RH, UK
\and
Space Telescope Science Institute, 3700 San Martin Drive, Baltimore, MD
21218-2463, USA
\and
Instituto de Astrof\'{\i}sica de Andaluc\'{\i}a (IAA-CSIC), Apartado
de Correos, 3004, E-18080 Granada, Spain
\and
NSSTC, SD-50, 320 Sparkman Dr., Huntsville, AL 35805, USA
\and
Istituto di Astrofisica Spaziale e Fisica Cosmica -- Sezione di Bologna, 
CNR, via Gobetti 101, I-40129 Bologna, Italy
\and
INAF -- Osservatorio Astronomico di Trieste, via G.B. Tiepolo 11, I-34131
Trieste, Italy
\and
Department of Physics and Astronomy, University of Canterbury,
Christchurch, New Zealand
\and
Joint Astronomy Centre, 660 N, A'ohoku Place, Hilo, Hawaii, HI 96720,
USA
\and
Department of Physics \& Astronomy, Clemson University, Clemson,
SC 29634-0978, USA
}

\date{\today}
   
\authorrunning{Vreeswijk, Ellison, Ledoux et al.}  

\titlerunning{\grb}
  
\abstract{We present photometry and spectroscopy of the afterglow of
  \grb. VLT spectra of the afterglow show damped \lya\ (DLA)
  absorption and low- and high-ionization lines at a redshift
  $z$=3.3718$\pm$0.0005. The inferred neutral hydrogen column density,
  log N(\ion{H}{i})=21.90$\pm$0.07, is larger than any (GRB- or QSO-)
  DLA \ion{H}{i} column density inferred directly from \lya\ in
  absorption. From the afterglow photometry, we derive a conservative
  upper limit to the host-galaxy extinction: A$_{\rm V}$$<$0.5 mag.
  The iron abundance is [Fe/H]=--1.47$\pm$0.11, while the metallicity
  of the gas as measured from sulphur is [S/H]=--1.26$\pm$0.20. We
  derive an upper limit on the H$_2$ molecular fraction of
  2$N$(H$_2$)/(2$N$(H$_2$)+$N$(\ion{H}{i}))$\lsim$10$^{-6}$.  In the
  \lya\ trough, a \lya\ emission line is detected, which corresponds
  to a star-formation rate (not corrected for dust extinction) of
  roughly 1 M\subsun\ yr$^{-1}$.  All these results are consistent
  with the host galaxy of \grb\ consisting of a low metallicity gas
  with a low dust content.  We detect fine-structure lines of silicon,
  \ion{Si}{ii}*, which have never been clearly detected in QSO-DLAs;
  this suggests that these lines are produced in the vicinity of the
  GRB explosion site. Under the assumption that these fine-structure
  levels are populated by particle collisions, we estimate the
  \ion{H}{i} volume density to be n$_{\ion{H}{i}}=10^2-10^4$
  cm$^{-3}$. HST/ACS imaging 4 months after the burst shows an
  extended AB(F606W)=28.0$\pm$0.3 mag object at a distance of
  0\farcs14 (1kpc) from the early afterglow location, which presumably
  is the host galaxy of \grb. } \maketitle

%
%________________________________________________________________

\section{Introduction}
\label{sec:introduction}

Damped \lya\ (DLA) absorbers, conventionally detected in Quasi-Stellar
Object (QSO) spectra, are absorption-line systems that have a column
density of N(\ion{H}{i}) $\ge$ 2$\times 10^{20}$ atoms cm$^{-2}$, as
determined from the damping wings of the \lya\ line
\citep[e.g.][]{1986ApJS...61..249W,1989ApJ...344..567T}.  DLA systems
are believed to contain the bulk of the neutral hydrogen at high
redshift and to be the gas reservoir from which the stars at the
present epoch are produced
\citep[e.g.][]{1987txra.symp..309W,1991ApJS...77....1L}. Numerous
high-resolution spectroscopic studies have extracted detailed
information about the metallicity \citep[e.g.][]{prochaska}, the
kinematics \citep{1997ApJ...487...73P,1998A&A...337...51L}, and the
dust and H$_2$ contents \citep{2000A&A...364L..26P,ledoux} of DLA
systems as a function of redshift.  Despite intensive searches, only a
handful of DLA counterparts have been detected so far
\citep[see][]{2002ApJ...574...51M}; linking DLA systems with galaxy
types has therefore proven difficult: some advocate large,
disk-forming galaxies
\citep[e.g.][]{1995ApJ...454..698W,1997ApJ...487...73P}, others
suggest they are faint, gas-rich dwarfs \citep{1998ApJ...495..647H}.
  
Gamma-ray burst (GRB) afterglows are, just as QSOs, bright and distant
sources. For instance, the spectacular GRB\,990123 was detected at the
9$^{\rm th}$ visual magnitude \citep{1999Natur.398..400A} while it was
located at $z$=1.6 \citep{1999Natur.398..389K,1999Sci...283.2075A}.
However, the afterglow brightness in general fades very rapidly in
time (roughly flux $\propto$ time$^{-1}$). The current afterglow
redshifts range from $z$=0.169 \citep{2003GCN..2020....1G} to $z$=4.5
\citep{2000A&A...364L..54A}.  Moreover, GRBs are associated with
massive-star formation: the discovery of a supernova (SN) spectrum
similar to that of SN1998\,bw \citep{1998Natur.395..670G} superimposed
on the GRB\,030329 afterglow spectrum \citep{stanek,hjorth030329}
provided strong evidence that at least some of the long-duration
($\gsim$ 2s) GRBs are caused by the core collapse of massive stars
\citep{1993ApJ...405..273W,1999ApJ...524..262M}.

The discovery of a damped \lya\ (DLA) absorption line at the burst
redshift in the spectra of several GRB afterglows
\citep{2001A&A...370..909J,fynbo926spectrum,hjorth020124} is
consistent with the massive-star progenitor scenario: they indicate a
high neutral hydrogen column density origin in the host galaxy,
presumably a star-forming region.  However, the signal-to-noise ratio
at the location of the DLA absorption line in the spectra is fairly
low in these cases, much lower than for typical QSO-DLAs.  We here
present afterglow spectra of the high-redshift \grb, which
unambiguously demonstrate a GRB-DLA, with a column density exceeding
that of any (QSO- or GRB-) DLA measured so far using \lya\ in
absorption. These spectra allow us to measure the metallicity of the
host environment and obtain an upper limit on the molecular fraction,
i.e.  measurements that are routinely performed for QSO-DLAs, but that
are still unique for GRB hosts.  Although the GRB-DLA sample is still
very small, we compare them with QSO-DLAs in two aspects: their
\ion{H}{i} column density and their metallicity.

\grb\ was detected on 23 March 2003 at 21:57 UT by HETE
\citep{2003GCN..1956....1G} with a fluence of 1.1$\times$10$^{-6}$
ergs cm$^{-2}$ (30-400 keV), and a duration of 26 seconds. Following
the HETE localization, the optical counterpart was discovered 7.6
hours after the burst at R.A. 11\hour06\min09\fs38, Decl.
--21\degree46\arcmin13\farcs3 (J2000) by \citet{2003GCN..1949....1G},
with a brightness of R=18.7. Our team reported a preliminary redshift
of $z$=3.372 \citep{2003GCN..1953....1V}, which is currently the third
highest redshift for a GRB
\citep{1998Natur.393...35K,2000A&A...364L..54A}.

This paper is organized as follows: in Sect. \ref{sec:observations},
we describe the data reduction of both the spectroscopic and imaging
observations. In Sect. \ref{sec:photometry}, we present the light
curves and infer an upper limit on the rest-frame optical extinction.
We measure the equivalent widths of the absorption lines and determine
the burst redshift in Sect. \ref{sec:redshift}. An \ion{H}{i} column
density model is fitted to the damped \lya\ line in Sect.
\ref{sec:column}, and we analyze the spectra in more detail in Sect.
\ref{sec:metallicityandh2} to derive the metallicity and an upper
limit on the molecular hydrogen (H$_2$) fraction. The detection of
\lya\ in emission is presented in Sect. \ref{sec:lya}, and we report
on the detection of the probable host galaxy of \grb\ in HST/ACS
imaging data in Sect. \ref{sec:hst}. In the final section, we close
with a general discussion of all these results.

\begin{table}[tp]
  \centering
  \caption[]{Log of UT4/FORS2 spectroscopic observations}\label{tab:spectroscopy}
  \null\vspace{-1.0cm}
  $$
  \begin{array}{cllccc}
    \hline
    \noalign{\smallskip}
    \rm UT \, date &
    \rm grism(filter) &
    \rm coverage &
    \rm \lambda / \Delta\lambda &
    \rm exptime &
    \rm seeing \\
    \rm March\,03 &
    &
    \rm (nm) &
    &
    \rm (min) &
    (\arcsec) \\
    \hline
    \rm 25.050 & \rm 300V        & 330-660  & 440  & 3\times10  & 1.1 \\
    \rm 25.077 & \rm 300I(OG590)  & 600-1100 & 660  & 3\times10  & 0.7 \\
    \rm 26.213 & \rm 1400V       & 456-586  & 2100 & 4\times30  & 0.8 \\
    \rm 26.306 & \rm 1200R(GG435) & 575-731  & 2140 & 4\times30  & 0.9 \\
    \hline
  \end{array}
  $$
\end{table}

\begin{table}[tbp]
  \centering
  \caption[]{Log of imaging observations}\label{tab:imaging}
  \null\vspace{-1.35cm}
  $$
  \begin{array}{rccccc}
    \hline
    \noalign{\smallskip}
    \rm UT \, date &
    \rm magnitude ^{\mathrm{a}} &
    \rm filter &
    \rm exptime &
    \rm seeing &
    \rm tel./instr.^{\mathrm{b}} \\
    \rm (2003) &
    &
    &
    \rm (min) &
    (\arcsec) &
    \\
    \hline
    \rm Mar\, 24.302 & 20.39 \pm 0.06 & \rm  B &  12 &  2.4 & \rm USNO\,1m  \\
    \rm Mar\, 24.310 & 19.68 \pm 0.05 & \rm  V &   8 &  2.5 & \rm USNO\,1m  \\
    \rm Mar\, 24.316 & 18.75 \pm 0.03 & \rm  R &   8 &  2.6 & \rm USNO\,1m  \\
    \rm Mar\, 24.323 & 18.20 \pm 0.04 & \rm  I &   8 &  2.4 & \rm USNO\,1m  \\
    \rm Mar\, 24.992 & 20.64 \pm 0.26 & \rm  R &8.33 &  1.1 & \rm CAHA\,2.2m  \\
    \rm Mar\, 25.027 & 20.56 \pm 0.08 & \rm  R &  54 &  2.0 & \rm Danish  \\
    \rm Mar\, 25.030 & 21.44 \pm 0.03 & \rm  V &   1 &  1.1 & \rm FORS2  \\
    \rm Mar\, 25.033 & 21.45 \pm 0.03 & \rm  V &   1 &  1.0 & \rm FORS2  \\
    \rm Mar\, 25.091 & 22.29 \pm 0.04 & \rm  B &   1 &  0.9 & \rm FORS2  \\
    \rm Mar\, 25.092 & 21.42 \pm 0.02 & \rm  V &   1 &  0.8 & \rm FORS2  \\
    \rm Mar\, 25.094 & 20.51 \pm 0.02 & \rm  R &   1 &  0.8 & \rm FORS2  \\
    \rm Mar\, 25.095 & 20.02 \pm 0.02 & \rm  I &   1 &  0.8 & \rm FORS2  \\
    \rm Mar\, 25.099 & 20.04 \pm 0.10 & \rm  I &  15 &  2.0 & \rm Danish  \\
    \rm Mar\, 25.129 & 20.57 \pm 0.03 & \rm  R &  25 &  2.0 & \rm Danish  \\
    \rm Mar\, 25.150 & 21.55 \pm 0.12 & \rm  V &  15 &  1.8 & \rm Danish  \\
    \rm Mar\, 25.250 & 20.83 \pm 0.13 & \rm  R & 130 &  3.0 & \rm SARA   \\
    \rm Mar\, 26.023 & 22.17 \pm 0.08 & \rm  V &  25 &  1.2 & \rm Danish  \\
    \rm Mar\, 26.041 & 21.27 \pm 0.05 & \rm  R &  25 &  1.0 & \rm Danish  \\
    \rm Mar\, 26.064 & 20.82 \pm 0.08 & \rm  I &  25 &  1.0 & \rm Danish  \\
    \rm Mar\, 26.107 & 19.86 \pm 0.04 & \rm  J &   6 &  0.7 & \rm NACO  \\
    \rm Mar\, 26.120 & 17.93 \pm 0.07 & \rm  K &2.25 &  0.7 & \rm NACO  \\
    \rm Mar\, 26.162 & 22.12 \pm 0.02 & \rm  V &   1 &  0.6 & \rm FORS2  \\
    \rm Mar\, 26.164 & 22.09 \pm 0.03 & \rm  V &   1 &  0.6 & \rm FORS2  \\
    \rm Mar\, 26.257 & 22.22 \pm 0.03 & \rm  V &   1 &  0.7 & \rm FORS2  \\
    \rm Mar\, 26.259 & 22.18 \pm 0.04 & \rm  V &   1 &  0.8 & \rm FORS2  \\
    \rm Mar\, 26.267 & 21.50 \pm 0.07 & \rm  R & 160 &  2.2 & \rm USNO\,1m  \\
    \rm Mar\, 26.306 & 18.13 \pm 0.18 & \rm  K &3.35 &  0.5 & \rm UKIRT  \\
    \rm Mar\, 26.351 & 22.32 \pm 0.04 & \rm  V &   3 &  0.8 & \rm FORS2  \\
    \rm Mar\, 26.351 & 19.38 \pm 0.05 & \rm  H &3.35 &  0.6 & \rm UKIRT  \\
    \rm Mar\, 26.353 & 21.43 \pm 0.02 & \rm  R &   3 &  0.8 & \rm FORS2  \\
    \rm Mar\, 26.356 & 20.90 \pm 0.03 & \rm  I &   3 &  0.8 & \rm FORS2  \\
    \rm Mar\, 26.379 & 20.09 \pm 0.03 & \rm  J &3.35 &  0.7 & \rm UKIRT  \\
    \rm Mar\, 27.041 & 22.01 \pm 0.14 & \rm  R &  30 &  1.0 & \rm Danish \\
    \rm Mar\, 27.156 & 22.94 \pm 0.09 & \rm  V &  24 &  0.7 & \rm Gemini\,S  \\
    \rm Mar\, 27.176 & 22.15 \pm 0.15 & \rm  R &22.5 &  0.6 & \rm Gemini\,S  \\
    \rm Mar\, 27.196 & 21.57 \pm 0.09 & \rm  I &22.5 &  0.6 & \rm Gemini\,S  \\
    \rm Mar\, 28.159 & 22.36 \pm 0.09 & \rm  R &  36 &  1.2 & \rm Danish \\
    \rm Mar\, 28.398 & 19.06 \pm 0.20 & \rm  K &   7 &  0.5 & \rm UKIRT  \\
    \rm Mar\, 28.455 & 20.34 \pm 0.08 & \rm  H &   7 &  0.7 & \rm UKIRT  \\
    \rm Apr\,  3.282 & 24.30 \pm 0.13 & \rm  R &  88 &  1.3 & \rm Gemini\,S  \\
    \rm Jul\,  5.979 & > 25.0 \,(3\sigma) & \rm  I &  15 &  0.7 & \rm FORS2 \\
    \rm Jul\,  5.993 & > 25.6 \,(3\sigma) & \rm  R &  15 &  0.8 & \rm FORS2 \\
    \rm Jul\, 20.958 & 28.0  \pm 0.3  & \rm V &  32 &      &  \rm HST \\
    \hline 
  \end{array}
  $$
  \begin{list}{}{}
  \item[$^{\mathrm{a}}$] The magnitudes have {\it not} been corrected
    for Galactic extinction, and the errors do {\it not} include the
    uncertainty in the absolute calibration (see text).
  \item[$^{\mathrm{b}}$] Combinations of telescopes and instruments:
    USNO Flagstaff Station 1.0m with 2k$\times$2k Tek CCD,
    0$\farcs$68/pixel; Calar Alto 2.2m and CAFOS with 1k$\times$1k
    SITe CCD, 0$\farcs$29/pixel; ESO/Danish 1.54m and DFOSC with
    2k$\times$4k EEV CCD, binned to 0$\farcs$78/pixel; Yepun and FORS2
    with two MIT CCDs of 4k$\times$2k, binned to 0$\farcs$25/pixel;
    Yepun and NACO with the Aladdin InSb 1k$\times$1k detector and S54
    camera, 0$\farcs$054/pixel; SARA 0.9m and Apogee Ap7 CCD camera
    with 512$\times$512 array, 0$\farcs$7/pixel; UKIRT and UFTI with a
    1k$\times$1k HgCdTe array, 0$\farcs$091/pixel; Gemini South and
    acquisition camera with 1K$\times$1K EEV CCD, binned to
    0$\farcs$23/pixel; HST/ACS and WFC detector with two 2k$\times$4k
    SITe CCDs, 0$\farcs$05/pixel.
  \end{list}
\end{table}

\section{Observations and data reduction}
\label{sec:observations}

The spectroscopic observations of \grb\ were performed with the Focal
Reducer Low Dispersion Spectrograph 2 (FORS2) at unit 4 (Yepun) of the
Very Large Telescope (VLT) at the European Southern Observatory (ESO)
at Paranal, Chile. The imaging observations were performed with a
variety of telescopes and instruments. Tables \ref{tab:spectroscopy}
and \ref{tab:imaging} show the spectroscopy and imaging observation
logs.

The images and 2-D spectra were bias-subtracted and flat-fielded in
the usual manner, mostly within IRAF\footnote{IRAF is distributed by
  the National Optical Astronomy Observatories, which are operated by
  the Association of Universities for Research in Astronomy, Inc.,
  under cooperative agreement with the National Science Foundation.}.
Following this, the spectra were cosmic-ray cleaned using the L.A.
Cosmic program written by \citet{2001PASP..113.1420V}. Each spectrum
was extracted separately, and wavelength-calibrated using an HeNeAr
lamp.  The error in the wavelength solution was of the order of
0.1\AA\ for the low resolution 300V and 300I grisms (with
$\lambda$/$\Delta \lambda$ of 440 and 660, respectively), and 0.03\AA\ 
for the intermediate resolution grisms 1400V and 1200R (with
$\lambda$/$\Delta \lambda$ $\sim$ 2100).  

Flux calibration was performed using the standard LTT3864, and the
slit losses were determined for each grism by fitting a Gaussian along
the spatial direction of the summed 2-D spectra, every 4 pixels across
the entire dispersion axis (i.e. summing 4 columns before performing
the fit).  The resulting Gaussian full width at half maximum (FWHM)
was then compared to the slit width to obtain the slit loss (i.e.  the
fraction of the surface underneath the Gaussian fit that is outside
the slit width) along the dispersion axis. The slit loss profile was
then fitted with a polynomial to correct the spectra.  All GRB spectra
were taken with a 1\arcsec\ slit, and slit losses were as high as 60\%
for the blue part of the 300V grism, while around 15\% for the other
grisms.  The standards were observed with 5\arcsec\ slits, and
therefore do not suffer from slit losses. We note that Yepun/FORS2
contains a linear atmospheric dispersion compensator (LADC) in the
light path, which minimizes any colour-differential slit losses up to
a zenith distance of 45\degr. However, the last three 1200R spectra
were taken at an airmass above this limit, and are therefore affected.

Finally, the first night's 300V+300I and second night's 1400V+1200R
spectra were combined into two full spectra and corrected for Galactic
extinction \citep{1998ApJ...500..525S}. The scaling between the blue
and red parts in the combining of the 300V with 300I spectra, and
1400V with 1200R, was determined from an overlapping region free of
lines, and amounted to 3.5\% of the continuum level between 300V and
300I, and 2\% for 1400V and 1200R, which can be explained by the
fading of the afterglow between the epochs at which the spectra were
taken. 

Comparison between the spectroscopy and imaging absolute
calibration shows that the spectroscopic flux is roughly 30\% below
the BVRI photometric measurements (i.e. not colour dependent) that
were taken around the same epoch. We have {\it not} corrected the
spectrum in Fig. \ref{spectrum} for this difference.

\begin{figure*}[t]
  \centering \includegraphics[width=18cm]{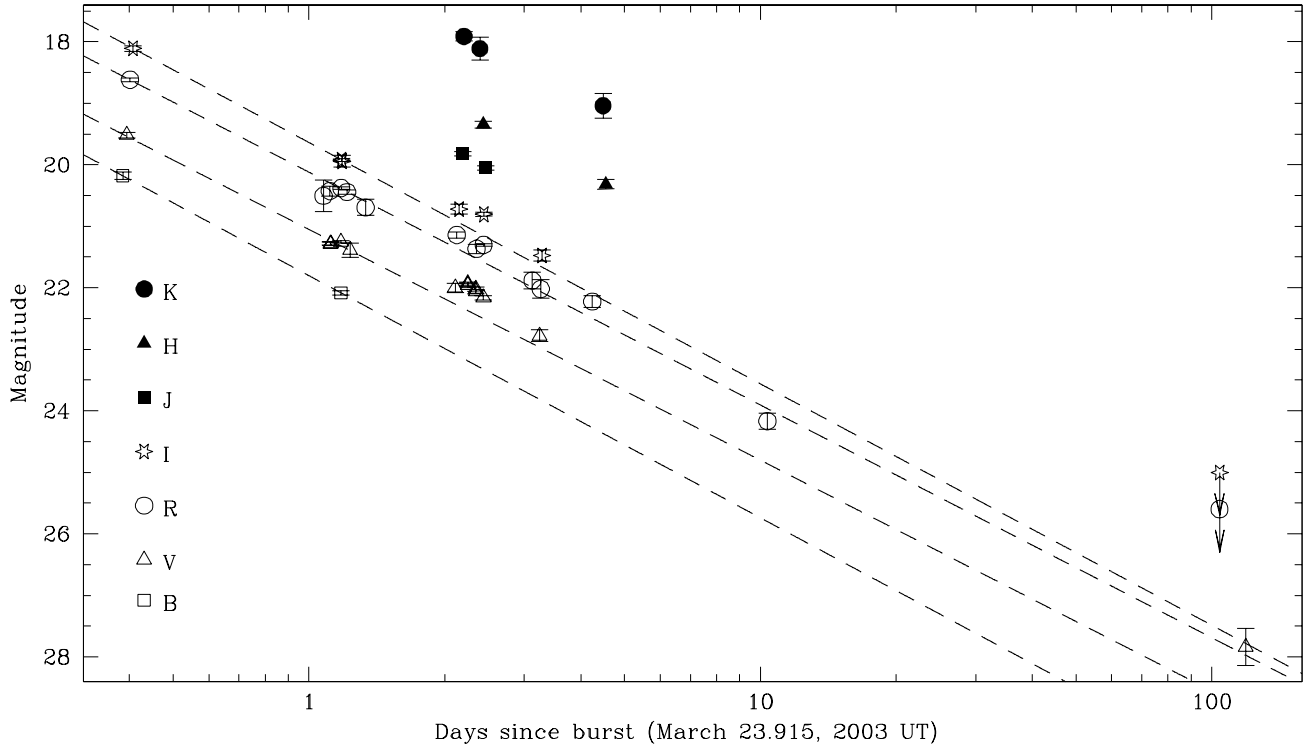}
  \caption{BVRIJHK light curves of \grb. The magnitudes have been
    corrected for the Galactic extinction of E$_{\rm B-V}$=0.049
    \citep{1998ApJ...500..525S}; note that in Table \ref{tab:imaging}
    they have {\it not} been corrected. The dashed lines are simple
    power law fits to the observations before day 2, and extrapolated
    to later epochs.
    \label{lightcurve}}
\end{figure*}

\section{Photometry}
\label{sec:photometry}

The magnitudes of the afterglow (see Table \ref{tab:imaging}) were
determined through aperture photometry, relative to a set of reference
stars. The aperture radius was set to the average FWHM of the point
spread function of stars on the image (column 5 in Table
\ref{tab:imaging}). The early July upper limits are 3$\sigma$, and
also for an aperture radius the size of the seeing disk. The reference
stars in the optical were tied to the absolute calibration provided by
\citet{2003GCN..1948....1H}. This was found to be consistent with our
own calibration on two different nights, but during which only one
standard field was observed.  For the near-infrared filters we used
the calibration provided by the 2MASS\footnote{see
  http://www.ipac.caltech.edu/2mass/}, as none of the nights on which
the near-infrared data were taken were photometric. The magnitude
errors listed in Table \ref{tab:imaging} are a combination of the
Poisson error of the afterglow measurement and the scatter in the
reference star magnitudes; they do {\it not} include the error in the
absolute calibration. For the optical bandpasses, we determined this
absolute error by calculating the average difference of two
calibrations on two different nights provided by
\citet{2003GCN..1948....1H} for several stars; we find the following
magnitude errors: 0.02 (B), 0.03 (V), 0.02 (R), and 0.12 (I). For the
infrared filters the absolute calibration error is provided by
\citet{2003AJ....126.1090C}, which amount to 0.02 mag for J, H and K.
No correction has been attempted for the fact that the observations
were taken with filters from different systems.  The magnitudes listed
in Table \ref{tab:imaging} have not yet been corrected for the
Galactic foreground absorption.

Fig. \ref{lightcurve} shows the light curves of \grb\ in several
wavelengths.  The light curves can be fit neither with a single power
law, nor with a broken power law.  This is clear from performing
simple power law fits to the early epochs in the optical bands, and
extrapolating these to later times.  These fits are shown by the
dashed lines in Fig.  \ref{lightcurve}.  The V band light curve shows
the clearest deviation in between day 2 and 3 after the burst.
However, after day 3 the observations are fairly well described again
by the extrapolation of the early slope, even the R band measurement
around day 10. The V band extrapolation underestimates the late-time
HST point by roughly a magnitude, which suggests that the afterglow
has become fainter than its host galaxy at this epoch.

Several afterglows have displayed deviations from the common smooth
power law decay, such as GRB\,970508
\citep[e.g.][]{1998ApJ...497L..13G}, GRB\,000301C
\citep[e.g.][]{2000A&A...359L..23M}, GRB\,021004
\citep[e.g.][]{2003AJ....125.2291H}, and GRB\,030329
\citep[e.g.][]{2003Natur.423..844P}. \citet{2000ApJ...544L..11G} have
suggested that a microlensing event caused the deviations in
GRB\,000301C.  In the context of the fireball model, the deviations in
the case of GRB\,021004 are interpreted as due to a variable external
density \citep{2002A&A...396L...5L}, and for GRB\,030329 they are
interpreted as due to refreshed shocks from the inner engine
\citep{granot030329}. We note that the cannonball model offers an
alternative explanation for these observations
\citep{2002A&A...388.1079D,2003ApJ...585L..15D,2003ApJ...594L..89D}.
For GRB\,030323, we only study the global properties of the light
curve, and compare them to the light curve decay and spectral slope
values as predicted by the fireball model
\citep[e.g.][]{1998ApJ...497L..17S,1999ApJ...519L..17S}, with the aim
of constraining the host-galaxy extinction.

As the late-time afterglow behaviour is not clear, we only use the
simple power law fits to the early optical data.  The inferred optical
temporal decay indices are consistent with one another, with an
average decay of $\alpha_{\rm opt}$=--1.56$\pm$0.03 (using the
convention: F$_{\nu}$(t) $\propto$ t$^{\alpha}\nu^{\beta}$). The
near-infrared slopes have similar values as the optical ones:
$\alpha_{\rm J}$ = --1.81$\pm$0.39, $\alpha_{\rm H}$ = --1.42$\pm$0.14
and $\alpha_{\rm K}$ = --1.46$\pm$0.31. However, these may be affected
by the ``bump'' in between day 2 and 3, if it is achromatic.

At several epochs after the burst, observations in at least two
filters were performed around the same epoch. This allows us to
construct broad-band spectral energy distributions (SEDs) and fit them
to obtain the optical to near-infrared spectral slopes at these
epochs.  Note that we discard the B and V bands, as these are
attenuated by the \lya\ line and forest absorption. For day after
burst 0.40, 1.18, 2.17, 2.44, 3.27, and 4.50, we obtain:
$\beta(0.40)$= --1.05$\pm$0.64 ($\chi^2_{\rm red}$=0), $\beta(1.18)$=
--0.83$\pm$0.61 ($\chi^2_{\rm red}$=0), $\beta(2.17)$=--1.09$\pm$0.07
($\chi^2_{\rm red}$=4.5), $\beta(2.44)$=--0.80$\pm$0.05 ($\chi^2_{\rm
  red}$=1.7), $\beta(3.27)$=--1.2$\pm$1.0 ($\chi^2_{\rm red}$=0), and
$\beta(4.50)$=--0.82$\pm$0.12 ($\chi^2_{\rm red}$=4.5). For the epochs
with $\chi^2_{\rm red}$=0, observations in only two filters are
available. Except for the fit value at day 2.17, these values are
consistent with being constant, and the weighted mean and its error is
$\beta_{\rm obs}$= --0.89$\pm$0.04.

We now compare these observed spectral slopes with the ones predicted
by the fireball model to obtain an estimate of the host-galaxy
extinction. An important assumption that we make in estimating the
optical extinction is that the intrinsic afterglow spectrum is a power
law, which is a consequence of the fireball theory for GRB afterglows.
The fireball theory has been quite successful in explaining the
observations \citep[but see][]{2002A&A...388.1079D}.  The predicted
spectral slope depends on the assumed circumburst density profile
being either constant or that of a stellar wind
\citep[see][]{2001ApJ...551..940L}, whether the light curve is in the
jet regime or not \citep[see][]{1999ApJ...519L..17S}, and whether the
cooling break \citep[see][]{1998ApJ...497L..17S} has already passed
the optical wavebands (cooling regime) or not.  Considering all these
possibilities, the predicted spectral slope ranges from $\beta_{\rm
  exp}$=($\alpha$+1)/2=--0.28$\pm$0.01 (wind or constant density
medium, post jet-break and non-cooling regime) to $\beta_{\rm
  exp}$=(2$\alpha$--1)/3=--1.37$\pm$0.01 (wind or constant density,
pre jet-break and cooling regime), with $\alpha$=--1.56$\pm$0.03.  We
have assumed that these relations between $\alpha$ and $\beta$ are
also valid for a power law index of the electron energy distribution,
$p<2$ \citep[but see][]{2001ApJ...558L.109D}.

\begin{figure}[tp]
  \centering \includegraphics[height=8.5cm,angle=-90]{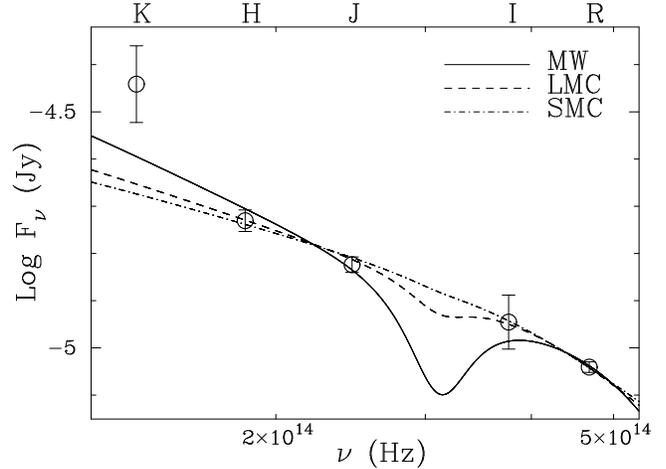}
  \caption{Milky Way (MW), Large- and Small Magellanic
    Cloud (LMC and SMC) extinction-curve fits to the spectral energy
    distribution (SED) of the afterglow at day 2.44 after the burst,
    assuming that the intrinsic SED is a power law with slope $\beta$=
    --0.28 (see text). These fits result in the conservative upper
    limits to the host-galaxy extinction: A$_{\rm V}$(MW)$<$0.50 mag,
    A$_{\rm V}$(LMC)$<$0.25 mag and A$_{\rm V}$(SMC)$<$0.16 mag.
    \label{fig:sed}}
\end{figure}

\begin{sidewaysfigure*}[tbp]
  \centering \includegraphics[width=25cm]{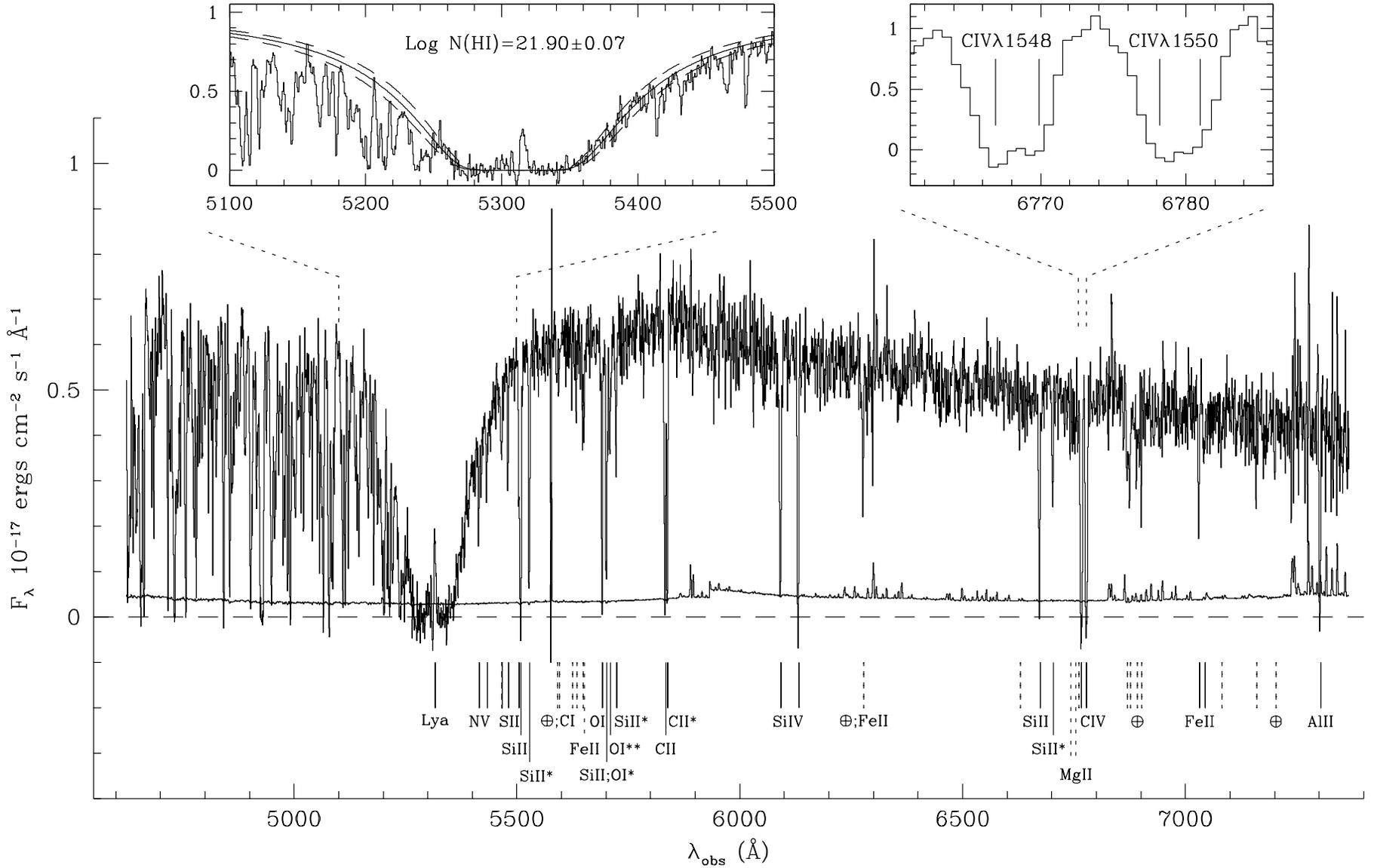}
  \caption{Combined 1400V+1200R spectrum of \grb, including
    the Poisson error spectrum. The inset on the left shows the
    normalized spectrum with the hydrogen column fit to the damped
    \lya\ line, including the 1$\sigma$ errors. This column density is
    currently the highest for any DLA system measured using \lya\ in
    absorption (see Fig. \ref{histo}). Redward of \lya\ we detect
    numerous metal-absorption lines (see Table \ref{tab:lines}), whose
    average redshift (in this high-resolution spectrum) is $z$=3.3716
    $\pm$ 0.0005. In the inset on the right, we zoom in on the
    \ion{C}{iv} doublet, which is split into separate components. In
    the \lya\ trough, \lya\ is detected in emission. Lines that could
    not be identified, atmospheric absorption lines (indicated with an
    $\oplus$), and lines belonging to the tentative absorber at
    $z$=1.41, are indicated with a dashed line instead of a solid one.
    \label{spectrum}}
\end{sidewaysfigure*}

All intrinsic spectral slopes shallower than the observed slope of
--0.89$\pm$0.04 leave some room for host-galaxy extinction
\citep[see][]{1998Natur.393...43R}, as any host-galaxy extinction
results in a steepening of the intrinsic slope.  We conservatively
take --0.28$\pm$0.01 to be the actual slope, to obtain an upper limit
on the host-galaxy extinction.  Using the extinction-curve fits of
\citet{1992ApJ...395..130P} for the Milky Way (MW), and the Large- and
Small Magellanic Clouds (LMC and SMC), we iteratively find A$_{\rm V}$
that fits best with the expected spectral slope of --0.28 (held fixed
in the fit) for the energy distribution at 2.44 days after the burst.
We find A$_{\rm V}$(MW)=0.50 mag ($\chi^2_{\rm red}$=1.5), A$_{\rm
  V}$(LMC)=0.25 mag ($\chi^2_{\rm red}$=1.8) and A$_{\rm V}$(SMC)=0.16
mag ($\chi^2_{\rm red}$=2.3). These fits are shown in Fig.
\ref{fig:sed}. We note that if we would have assumed that the
light-curve break occurred after day 1.4, which is likely, then the
expected spectral slope would be $\beta_{\rm
  exp}$=(2$\alpha$+1)/3=--0.71, and the extinction values would
decrease to A$_{\rm V}$(MW)=0.09 mag ($\chi^2_{\rm red}$=1.3), A$_{\rm
  V}$(LMC)=0.04 mag ($\chi^2_{\rm red}$=1.4) and A$_{\rm V}$(SMC)=0.02
mag ($\chi^2_{\rm red}$=1.4). When we do not fix the intrinsic slope
at a particular value (but still assume that the intrinsic spectrum is
a power law), we find spectral slopes ranging from --0.80 to --0.67,
and A$_{\rm V}$ from 0 to 0.12 mag.  Therefore, a very conservative
upper limit on the host galaxy extinction is: A$_{\rm V}$$<$0.50 mag.

\section{Absorption-line measurements and redshift}
\label{sec:redshift}

The combined 1400V+1200R spectrum of \grb, corrected for Galactic
extinction, is shown in Fig. \ref{spectrum}. The most obvious feature
is the very broad absorption line around 5300\AA, which can be
identified as \lya.  Redward of \lya\ several metal absorption lines
are detected, and to the blue the intervening \lya\ forest is present.
We also show the 1$\sigma$ Poisson error spectrum.
  
After normalization of the spectrum with a high-order (25) polynomial,
we measured all possible lines with {\it splot} in IRAF, summing the
equivalent width (EW) of the individual pixels, and determined the
line center.  In case two lines were clearly blended, we used {\it
  splot} to deblend, using Gaussian line shapes, and forcing a single
FWHM for both lines. The EWs and centers of the lines above 5$\sigma$
significance are tabulated in Table \ref{tab:lines} for both the low-
(lr) and high-resolution (hr) spectrum, along with the error in the
EW, the line identification and the line redshift. These line
redshifts are determined from the high-resolution spectrum, unless the
line is not covered by this spectrum (i.e.  above 7310\AA\ and below
4560\AA\ -- see Table \ref{tab:spectroscopy}). The error in the EW is
determined by: $\rm \Delta EW\, = \Delta \lambda \, \sqrt{\Sigma
  \sigma_i^2} $, where the error spectrum ($\sigma$) is normalized by
the high-order polynomial continuum fit of the object spectrum, and is
summed over the same pixels (i) that were used to measure the line EW.
$\Delta \lambda$ is the number of \AA\ per pixel, which is 3.2 and
0.64 \AA/pixel for the combined 300V+I and 1400V+1200R spectra,
respectively.
 
Using all lines that are detected above 5$\sigma$ significance and
that could be identified, we find $z$(lr)=3.3728$\pm$0.0011 and
$z$(hr)=3.3716$\pm$0.0005, for the low- (lr) and high-resolution (hr)
spectrum, respectively. We adopt the weighted mean of these values as
the redshift of \grb: $z$=3.3718$\pm$0.0005. There is no doubt that
this is the redshift of \grb\ and not that of a chance foreground
galaxy, since otherwise \lya\ forest lines redward of the DLA line
would have been detected. We find that the strongest lines in the red
part of the high-resolution spectrum, \ion{C}{iv}\lam\lam 1548, 1550, are
split into separate components (see the inset of Fig.  \ref{spectrum})
with a velocity difference of 130$\pm$60 km s$^{-1}$ (where the error
on the wavelength determination of the lines in the blend is 1\AA).
Such a velocity spread is consistent with the absorption taking place
in separate regions in the host galaxy.

Several lines that are detected above 5$\sigma$ can not be identified,
some of which correspond to significant lines in the standard star
spectra. Most of these are imprinted on the spectrum by the Earth's
atmosphere. Several lines not belonging to the GRB host galaxy can in
principle be identified with \ion{Fe}{ii} \lam\lam 2344.2, 2600.1
(although an atmospheric line is detected at the latter wavelength)
and \ion{Mg}{ii} \lam\lam 2796.3, 2803.5, all around $z$=1.41. However,
the oscillator strength of \ion{Fe}{ii} \lam 2382 is three times larger
than that of the detected \ion{Fe}{ii} \lam 2344, but this line is not
detected. Moreover, both \ion{Fe}{ii} lines are stronger than
\ion{Mg}{ii}, which is usually not observed
\citep[e.g.][]{1985A&A...145...59B,1992ApJS...80....1S}.  Therefore,
we consider the existence of this foreground absorption system to be
highly uncertain.
  
The optical/UV flash of the GRB is expected to alter its immediate
environment, possibly leading to a change in absorption-line strengths
as a function of time
\citep[see][]{1998ApJ...503L.135P,2001ApJ...546..672V,2003ApJ...585..775P}.
Comparing the low- and high-resolution equivalent widths in Table
\ref{tab:lines} shows that none of the lines detected in both spectra
are significantly varying (3$\sigma$).

\begin{table}[tbp]
  \centering
  \caption[]{Lines detected above 5$\sigma$ in the low- (lr) and high-resolution (hr) spectra. \label{tab:lines}}
  \null\vspace{-1.35cm}
  $$
  \begin{array}{ccccrc}
    \hline
    \noalign{\smallskip}
    \rm \lambda(lr)  &
    \rm EW_{\rm rest}(lr) &
    \rm \lambda(hr) &
    \rm EW_{\rm rest}(hr) &
    \rm ID^{\mathrm{b,c}} &
    \rm z \\
    \hline
 4484 &   9 (1)^{\mathrm{a}} & &                 &\rm             $\lyb$\,\lambda 1025.7  & 3.3715\\
 5317 &               & 5316.9  &                 &       $\lya$\,\rm em.\,\lambda 1215.6  & 3.3736\\
      &               & 5415.8  & 0.31 (3)      &\rm         \ion{N}{v}\,\lambda 1238.8  & 3.3717\\
      &               & 5433.5  & 0.20 (2)      &\rm         \ion{N}{v}\,\lambda 1242.8  & 3.3720\\
      &               & 5465.4  & 0.15 (2)      &                             &       \\
      &               & 5467.4  & 0.21 (3)      &\rm        \ion{S}{ii}\,\lambda 1250.5  & 3.3719\\ 
 5483 & 0.44 (6)      & 5481.1  & 0.29 (2)      &\rm        \ion{S}{ii}\,\lambda 1253.8  & 3.3716\\
 5511 & 1.07 (5)      & 5505.0  & 0.33 (2)      &\rm        \ion{S}{ii}\,\lambda 1259.5  & 3.3707\\
      &               & 5509.7  & 1.00 (2)      &\rm       \ion{Si}{ii}\,\lambda 1260.4  & 3.3713\\
      &               &         &                 &\rm  or\, \ion{Fe}{ii}\,\lambda 1260.5  & 3.3709\\
 5530 & 0.76 (5)      & 5529.2  & 0.75 (2)      &\rm   \ion{Si}{ii}^{*}\,\lambda 1264.7  & 3.3718\\
      &               & 5591.5  & 0.11 (2)      &\oplus,\rm \ion{C}{i}\,\lambda 1280.1 ? & 3.3679\\
      &               & 5595.7  & 0.10 (2)      &                             &       \\   
      &               & 5624.4  & 0.11 (2)      &                             &       \\  
 5637 & 0.33 (5)      & 5634.7  & 0.16 (2)      &                             &       \\  
      &               & 5649.2  & 0.16 (1)      &                             &       \\
      &               & 5652.0  & 0.15 (1)      &\rm      \ion{Fe}{ii}\,\lambda 2344.2 ? & 1.4110\\
 5692 & 0.75 (6)      & 5692.5  & 0.69 (2)      &\rm         \ion{O}{i}\,\lambda 1302.1  & 3.3716\\
 5706 & 1.08 (6)      & 5702.4  & 0.70 (2)      &\rm       \ion{Si}{ii}\,\lambda 1304.3  & 3.3718\\
      &               &         &               &\rm and\,\ion{O}{i}^{*}\,\lambda 1304.8 ?& 3.3702\\
 5706 & 1.08 (6)      & 5709.7  & 0.32 (2)      &\rm    \ion{O}{i}^{**}\,\lambda 1306.0 ? & 3.3718\\
 5727 & 0.26 (5)      & 5723.9  & 0.29 (2)      &\rm   \ion{Si}{ii}^{*}\,\lambda 1309.2  & 3.3718\\
 5838 & 1.71 (7)      & 5833.7  & 0.84 (3)      &\rm        \ion{C}{ii}\,\lambda 1334.5  & 3.3713\\
      &               & 5838.9  & 0.79 (2)      &\rm    \ion{C}{ii}^{*}\,\lambda 1335.7  & 3.3714\\
 6095 & 1.19 (6)      & 6092.9  & 1.08 (4)      &\rm       \ion{Si}{iv}\,\lambda 1393.7  & 3.3716\\
 6135 & 0.98 (6)      & 6132.4  & 1.15 (4)      &\rm       \ion{Si}{iv}\,\lambda 1402.7  & 3.3716\\
 6280 & 0.51 (3)      & 6278.0  & 0.42 (3)      &\oplus,\rm \ion{Fe}{ii}\lambda 2600.1 ? & 1.4145\\
      &               & 6629.2  & 0.14 (2)      &                             &       \\
 6676 & 0.69 (4)      & 6673.8  & 0.75 (3)      &\rm       \ion{Si}{ii}\,\lambda 1526.7  & 3.3714\\
 6706 & 0.54 (4)      & 6703.7  & 0.34 (3)      &\rm   \ion{Si}{ii}^{*}\,\lambda 1533.4  & 3.3717\\
      &               & 6743.0  & 0.14 (2)      &\rm      \ion{Mg}{ii}\,\lambda 2796.3 ? & 1.4114\\
      &               & 6754.2  & 0.13 (2)      &\rm      \ion{Mg}{ii}\,\lambda 2803.5 ? & 1.4092\\
      &               & 6760.6  & 0.14 (3)      &                             &       \\
 6770 & 1.42 (4)      & 6766.9  & 0.99 (3)      &\rm        \ion{C}{iv}\,\lambda 1548.2  & 3.3708\\
      &               & 6769.9  & 0.51 (3)      &\rm        \ion{C}{iv}\,\lambda 1548.2  & 3.3728\\
 6782 & 1.28 (4)      & 6778.2  & 0.84 (3)      &\rm        \ion{C}{iv}\,\lambda 1550.7  & 3.3708\\
      &               & 6781.0  & 0.53 (3)      &\rm        \ion{C}{iv}\,\lambda 1550.7  & 3.3727\\
 6875 &               & 6870.4  &                 &\oplus                    &       \\
      &               & 6876.5  &                 &\oplus                    &       \\
 6900 &               & 6891.9  &                 &\oplus                    &       \\
      &               & 6902.4  &                 &\oplus                    &       \\
 7034 & 0.51 (4)      & 7031.7  & 0.48 (3)      &\rm       \ion{Fe}{ii}\,\lambda 1608.4  & 3.3717\\
      &               & 7043.6  & 0.17 (3)      &\rm       \ion{Fe}{ii}\,\lambda 1611.2  & 3.3716\\
      &               & 7082.0  & 0.15 (3)      &                             &       \\
      &               & 7160.1  & 0.20 (3)      &                             &       \\
      &               & 7202.7  & 0.17 (3)      &\oplus                    &       \\
 7307 & 1.08 (5)      & 7303.5  & 0.99 (7)      &\rm       \ion{Al}{ii}\,\lambda 1670.7  & 3.3713\\
 7907 & 0.58 (5)      &         &                 &\rm       \ion{Si}{ii}\,\lambda 1808.0  & 3.3734\\
 8112 & 0.45 (5)      &         &                 &\rm      \ion{Al}{iii}\,\lambda 1854.7  & 3.3738\\
 8145 & 0.23 (4)      &         &                 &\rm      \ion{Al}{iii}\,\lambda 1862.7  & 3.3724\\
 8237 & 0.34 (4)      &         &                 &                             &       \\
 8802 & 0.55 (6)      &         &                 &                             &       \\
 8859 & 0.82 (9)      &         &                 &\rm       \ion{Zn}{ii}\,\lambda 2026.1  & 3.3725\\
 8970 & 0.73 (11)     &         &                 &\oplus                    &       \\
 9019 & 0.77 (5)      &         &                 &\oplus,\rm \ion{Zn}{ii}\,\lambda 2062.6  & 3.3726\\
  \end{array}
  $$
  \null\vspace{-0.5cm}
  \begin{list}{}{}
  \item[$^{\mathrm{a}}$] \lyb\ is located in the \lya\ forest, which
    causes it to be blended with forest lines, and the continuum
    placement is highly uncertain, resulting in a large EW error.
  \item[$^{\mathrm{b}}$] The lines at \lam\lam 5652.0, 6278.0, 6743.0,
    and 6754.2 can in principle be identified with \ion{Fe}{ii} and
    \ion{Mg}{ii} at $z$=1.41; however, strong absorption is then
    expected from \ion{Fe}{ii} \lam 2382, which is not detected.
  \item[$^{\mathrm{c}}$] The lines for which significant absorption
    was also detected in the standard star spectrum are marked with a
    $\oplus$.
  \end{list}
\end{table}

\section{\ion{H}{i} column density}
\label{sec:column}

We have fitted a power law continuum to the high-resolution spectrum
over the wavelength range 5870--7000 \AA\ and determined a power law
slope (in $F_{\nu}$) of --0.76$\pm$0.09 (fitting only the first 1200R
spectrum, as the other 1200R spectra suffer from colour-dependent slit
losses).  This value is in agreement with the red slope of the 300I
spectrum and with the slope of the photometry measurements (see Sect.
\ref{sec:photometry}).  We used an extrapolation of this power law to
blue wavelengths in order to normalize the entire spectrum. The
resulting average flux decrement in the \lya\ forest between \lyb\ and
\lya: $D_A = 1-F_{\nu}(\rm observed)/F_{\nu}(\rm intrinsic)$
\citep{1982ApJ...255...11O}, that we obtain is $D_A$=0.44 $\pm$0.04.
This decrement, which is due to intervening hydrogen systems, is
consistent with that observed in QSO lines of sight at the redshift of
\grb\ \citep{1993A&A...268...86C}.

\begin{figure}[bp]
  \centering \includegraphics[width=8.5cm]{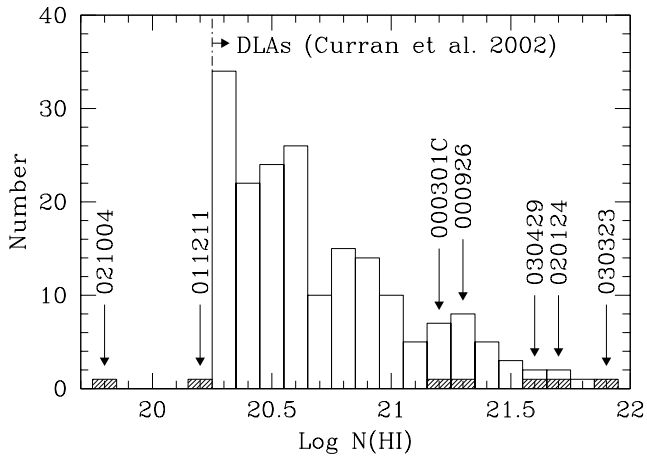}
  \caption{Histogram of the column densities of DLA systems
    measured through the damping wings of \lya\ discovered in the
    spectrum of a background QSO \citep[compilation taken
    from][]{2002PASA...19..455C}. The shaded histogram shows
    measurements in GRBs for which the redshift was large enough to
    detect \lya. Out of 7 GRBs, 5 show neutral hydrogen column
    densities above the DLA definition of 2$\times$10$^{20}$ atoms
    cm$^{-2}$ (log N(\ion{H}{i})=20.3).  The host of \grb\ contains a
    column density larger than in any observed (GRB- or QSO-) DLA
    system.
\label{histo}}
\end{figure}

A fit to the strong Ly$\alpha$ absorption (using the DIPSO package
within Starlink) yields log N(\ion{H}{i})=21.90$\pm$0.07. This fit is
shown in the inset of Fig.  \ref{spectrum}. This is the 7th GRB for
which a neutral hydrogen column has been determined from the afterglow
spectrum, and \grb\ happens to have the highest \ion{H}{i} column
density measured so far.  Fig.  \ref{histo} shows a comparison of the
\ion{H}{i} column density distribution of QSO-DLAs \citep[taken from
the compilation of][]{2002PASA...19..455C} and GRB-DLAs \citep[][
Jakobsson et al.  2004, in prep., and this
paper]{2001A&A...370..909J,fynbo926spectrum,hjorth020124}.  For
completeness, we also show the two GRBs for which \lya\ was detected
but which do not qualify as a DLA system: GRB\,011211 (Vreeswijk et
al. 2004, in prep.) and GRB\,021004 \citep{2002A&A...396L..21M}.  It
is quite striking that out of 7 GRB afterglows for which Ly$\alpha$
was red-shifted into the observable spectrum, 5 show evidence for a
high column density DLA system. This clearly demonstrates that GRBs
explode in either galaxies, or regions within galaxies with high
neutral hydrogen column densities. The \ion{H}{i} gas responsible for
these large columns could be related to the site of the GRB explosion,
e.g. part of the massive-star forming region in which the GRB
occurred, but could also be gas that is not associated with the GRB,
further away in the host galaxy. We performed a Kolgomorov-Smirnov
(KS) test \citep[e.g.][]{numrec} to estimate that the probability that
both samples are drawn from the same parent distribution is 0.0006.
Moreover, in this comparison with QSO-DLAs, the GRB-DLA \ion{H}{i}
column densities are in fact lower limits as the GRB itself occurs
within the galaxy that is associated with the DLA system; if the GRB
sightlines would have been probed with background QSOs, their column
densities would have been on average a factor of two larger, which
would shift the GRB column densities in Fig.  \ref{histo} by 0.3 dex
upward.  However, GRB\,011211 would then move into the GRB-DLA sample
resulting in a only a slight decrease in the above-mentioned KS
probability.

\section{Metallicity and H$_2$ content}
\label{sec:metallicityandh2}

Although there are many metal lines observed in the spectrum of this
GRB, most of them are saturated in the intermediate resolution
spectrum.  We have identified only 2 sets of lines as potentially
unsaturated, based on their small ($<0.4$\AA) rest-frame equivalent
widths (EWs).  These are the \ion{S}{ii} $\lambda1250,1253,1259$
triplet and \ion{Fe}{ii} $\lambda 1611$.  Two of the \ion{S}{ii} lines
($\lambda=1250,1259$) show signs of blending, evidenced by a weak
component that broadens the $\lambda 1250$ line in its blue wing, and
a strong interloper redward of the $\lambda 1259$ line.  \ion{S}{ii}
$\lambda 1253$ appears as an unresolved single component. We measured
the observed EWs of \ion{S}{ii} $\lambda 1253$ and \ion{Fe}{ii}
$\lambda 1611$ to be 1.25\AA\ and 0.72\AA\ (EW$_{\rm rest}$= 0.29\AA\ 
and 0.16\AA), respectively.  In the optically thin limit, these
correspond to the column densities: log N(\ion{S}{ii})=15.3 and log
N(\ion{Fe}{ii})=15.7, and abundances [S/H]=--1.8 and [Fe/H]=--1.7.  In
this conversion from column density to abundance, we assumed the Solar
values from \citet{1998SSRv...85..161G}, and no correction for
ionization; i.e. we assumed that the column densities of \ion{S}{ii}
and \ion{Fe}{ii} are equal to the total column densities of S and Fe,
as the singly ionized state of both of these elements should be the
dominant one in a region with such a high \ion{H}{i} column density.
This has been motivated theoretically for QSO-DLAs
\citep{1995MNRAS.276..268V,2001ApJ...557.1007V}; we here assume that
ionization corrections are also negligible in GRB-DLAs.

\begin{figure}[tp]
  \centering \includegraphics[width=8.5cm]{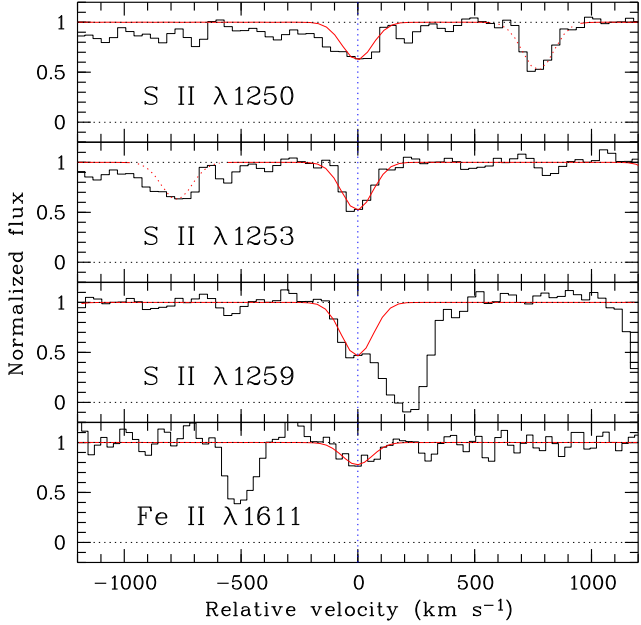}
  \caption{Simultaneous one-component fit to low-ionization
    metal lines with little saturation.\label{fit}}
\end{figure}

In a second step, we performed a simultaneous one-component fit to the
lines \ion{S}{ii} $\lambda1250,1253,1259$ and \ion{Fe}{ii} $\lambda
1611$, taking into account line blending and a range of broadening
parameters. Fig. \ref{fit} shows the resulting fits. We find log
N(\ion{S}{ii})=15.84$\pm$0.19 and log N(\ion{Fe}{ii})=15.93$\pm$0.08,
together with a (turbulent) broadening parameter b=35$\pm$10 km
s$^{-1}$. Comparing these column densities with those from the
optically thin limit approximation shows that the \ion{S}{ii} $\lambda
1253$ and \ion{Fe}{ii} $\lambda 1611$ lines are slightly saturated.
These measured column densities correspond to the abundances:
[S/H]=--1.26$\pm$0.20 and [Fe/H]=--1.47$\pm$0.11.  Therefore, although
the N(\ion{Fe}{ii}) is large \citep[see also][]{2003ApJ...585..638S},
the [Fe/H] is only marginally higher than that of QSO-DLAs at this
redshift: the mean [Fe/H] of 26 QSO-DLAs with 3.0$<z<$3.5 from
\citet{prochaska} is --1.83, with a scatter of 0.35.

Since essentially all DLA systems observed toward QSOs have [S/Fe]
greater than zero \citep[see][ for a recent compilation and
discussion]{lopezellison}, our measurement of [S/Fe]=0.21$\pm$0.23 in
\grb\ does not represent a very stringent constraint on a possible
$\alpha$-element overabundance.  Moreover, as iron is a known
dust-depleted element, there could be a small correction to this ratio
due to dust depletion. Interestingly, a tendency toward high values of
[Si/Fe]\footnote{Observationally, [S/Si]=0 in gaseous absorbers when
  there is neither dust depletion nor ionization effects.} has been
found in other GRBs \citep{2003ApJ...585..638S}, as expected in cases
where massive-star formation has recently deposited metals into the
ISM.

We have examined the \grb\ spectra for presence of H$_2$ absorption
lines, but these are not detected \citep[for a list of lines and their
oscillator strengths, see][]{1976ApJ...204....1M}. The location of
possible H$_2$ lines at z$_{\rm abs}$=3.3716 is actually observed for
the L=0 to 3 Lyman bands of H$_2$.  Of these, only the expected
location of the L=2 band is clear of blending with \lya\ forest lines.
Because of the low resolution of the spectra, two ranges of possible
broadening parameters b$_{\rm H_2}$ (with b$_{\rm H_2}\lsim$ b$_{\rm
  metals}$) were considered to perform trials of Voigt-profile fitting
of both the J=0 and 1 lines (namely: H$_2$ L2-0 R(0), L2-0 R(1) and
L2-0 P(1)). We find the following upper limits: (1) for the range 10
km s$^{-1}$ $<$ b$_{\rm H_2}\lsim$ 50 km s$^{-1}$: log N(J=0)$<$14.5
and log N(J=1)$<$15.5, and (2) for the range 1 km s$^{-1}\lsim$
b$_{\rm H_2}<$ 10 km s$^{-1}$: log N(J=0)$<$14 and log N(J=1)$<$18.
Therefore, strictly speaking the derived upper limit on the mean
molecular fraction of the system (i.e. GRB environment + host galaxy)
is: f $\equiv$ 2N(H$_2$)/(2N(H$_2$)+N(\ion{H}{i})) $<$
2$\times$10$^{18}$/(2$\times$10$^{18}$+7.9$\times$10$^{21}$) =
2.5$\times$10$^{-4}$ with N(\ion{H}{i})=7.9$\times$10$^{21}$ cm$^{-2}$
(i.e. the above case 2).  However, under the assumption that
N(J=1)$\lsim$10$\times$N(J=0), as observed in H$_2$-detected QSO-DLAs
\citep{ledoux} and in the Magellanic Clouds
\citep{2002ApJ...566..857T}, i.e. taking log N(J=0)=14.5 and log
N(J=1)=15.5 (which is actually the above case 1), f should be less
than or of the order of 10$^{-6}$.  Although our spectra have a lower
spectral resolution than those normally used to study DLA systems
along QSO lines of sight, the large \ion{H}{i} column density in \grb\ 
allowed us to obtain an upper limit which is similar to the limits
found in QSO-DLAs.  We also examined the \grb\ spectra for presence of
absorption lines from vibrationally excited molecular hydrogen
predicted by \citet{2002ApJ...569..780D}, but these are also not
detected.

As shown by \citet{ledoux}, the lack of H$_2$ molecules in DLA systems
is mainly due to the low metallicity of the gas in addition to its
particular physical conditions (density, temperature, UV flux). In
particular, H$_2$ is usually not detected whenever the metallicity
[X/H]$<$--1. In \grb, the sulphur metallicity, [S/H]=--1.26$\pm$0.20,
is low enough to explain the lack of H$_2$. An alternative explanation
is that H$_2$ close to the GRB has been dissociated by the strong
UV/X-ray emission; however such an emission would also ionize a large
fraction of the gas with which H$_2$ molecules are associated
\citep[see][]{2002ApJ...569..780D}.

\begin{figure}[bp]
  \centering \includegraphics[width=8.5cm]{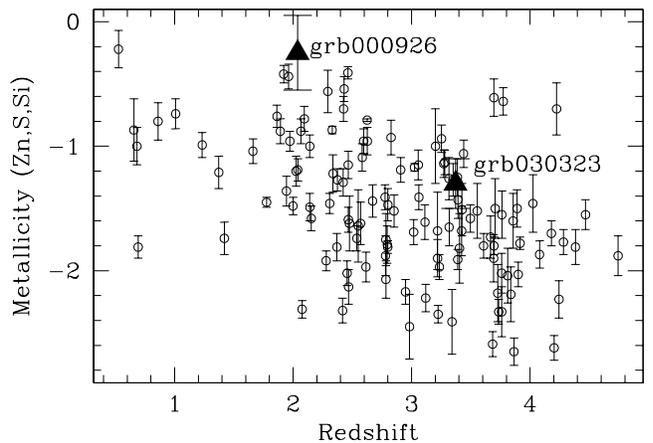}
  \caption{Comparison of the metallicities of a sample
    of QSO-DLAs, taken from \citet{prochaska} (open circles), with the
    two GRBs for which a metallicity has been determined (solid
    triangles): GRB\,000926 and \grb\ (this paper). The GRB hosts are
    located at the metal-rich end of the QSO-DLA distribution.
      \label{fig:metalz}}
\end{figure}

A dust depletion factor (i.e. the abundance difference between a
dust-depleted element such as Fe or Cr and a non-depleted element such
as Zn or S) of 0.2 dex at a metallicity of --1.26 (cf.
[Fe/H]=--1.47$\pm$0.11) is also consistent with measurements in
QSO-DLAs \citep[see Fig. 12 of][]{ledoux}.  However, this result is
different from the analysis of three GRB host galaxies by
\citet{2003ApJ...585..638S}, who find that the GRB host dust depletion
is much larger than it is in QSO-DLAs.

In Fig. \ref{fig:metalz}, we compare the metallicities (from Zn, S or
Si) of a sample of QSO-DLAs taken from \citet{prochaska}, with the
GRB-DLAs for which a metallicity has been determined: GRB\,000926 and
\grb\ (this paper). For GRB\,000926, we have adopted the value
[Zn/H]=--0.25 of \citet{castro926}, which is consistent with the
curve-of-growth analysis value -- [Zn/H]=--0.13 -- of
\citet{2003ApJ...585..638S}. Since the neutral hydrogen column density
determination for GRB\,000926 is not secure \citep{fynbo926spectrum},
we assume an error of 0.3 dex.  Although only two GRBs have measured
metallicities, Fig.  \ref{fig:metalz} suggests that GRB host galaxies
are more metal rich than QSO-DLAs.  \citet{2003ApJ...585..638S}
already pointed out large Zn column densities in three GRB host
galaxies (for which only GRB\,000926 has a measured \ion{H}{i} column
density) with respect to QSO-DLAs, while they found the Fe column
densities to be similar to those of QSO-DLAs.  Hence, [Zn/Fe], a
measure of the amount of dust depletion, is very large in their sample
of GRB hosts with respect to QSO-DLAs, suggestive of a high dust
content.  Although we do not have an estimate of the Zn column
density, the quantity [S/Fe] is a similar measure.  For \grb, we find
[S/Fe]=0.2, while \citet{2003ApJS..147..227P} find $<$[S/Fe]$>$=0.4
(based on three systems in their sample for which this quantity is not
an upper or lower limit).

Using the measured metallicity and \ion{H}{i} column density, we can
check the low optical extinction that we inferred from the afterglow
photometry.  Following \citet{2002ApJ...566...68P}, we assume that
A$_{\rm V}$=R$_{\rm V} \kappa$ N(\ion{H}{i})$_{\rm
  host}$/4.9$\times$10$^{21}$, where R$_{\rm V}\equiv$A$_{\rm
  V}$/E(B-V) is the total-to-selective extinction; R$_{\rm
  V}$(MW)=3.1, R$_{\rm V}$(LMC)=3.2, and R$_{\rm V}$(SMC)=2.9
\citep[see][]{1992ApJ...395..130P}. The dust-to-gas ratio,
$\kappa$=10$^{\rm [X/H]}$(1-10$^{\rm [Fe/X]}$), corresponds to the
dust-to-gas ratio of the dust responsible for the extinction. The
value (4.9$\pm$0.3)$\times$10$^{21}$ cm$^{-2}$ mag$^{-1}$ corresponds
to the Galactic value for N(\ion{H}{i})/E(B-V)
\citep{1994ApJ...427..274D}; for the LMC and SMC, we assume the values
(2.0$\pm$0.5)$\times$10$^{22}$ cm$^{-2}$ mag$^{-1}$
\citep{1982A&A...107..247K} and (4.4$\pm$0.7)$\times$10$^{22}$
cm$^{-2}$ mag$^{-1}$ \citep{1985A&A...149..330B}, respectively.  With
[S/H]=--1.26 and [Fe/S]=--0.21, we find $\kappa$=0.02, and A$_{\rm
  V}$(MW)=0.08 mag, A$_{\rm V}$(LMC)=0.02 mag, and A$_{\rm
  V}$(SMC)=0.01 mag.  These values are all consistent with the upper
limits derived from the afterglow photometry (see Sect.
\ref{sec:photometry}).

As can be seen in Fig. \ref{spectrum}, fine-structure lines of both
\ion{C}{ii}* and \ion{Si}{ii}* are detected. The \ion{Si}{ii}* lines
have never been clearly detected in QSO-DLAs, which suggests that
their origin is associated with \grb, or that they can be found only
in regions with very high neutral gas densities.  \ion{Si}{ii}* \lam
1264 has also been observed along the GRB\,010222 sightline (I.
Salamanca, private communication). The population of the
fine-structure levels is a function of the density of the absorbing
medium and the ambient photon-flux intensity
\citep{1968ApJ...152..701B}. Using the calculations of
\citet{2002MNRAS.329..135S}, we can make a rough estimate of the
\ion{H}{i} volume density using the two un-saturated \ion{Si}{ii}*
lines \lam\lam 1309,1533 (the \ion{C}{ii}* and \ion{Si}{ii}* \lam 1264 lines
{\it are} saturated), for which we measure log
N(\ion{Si}{ii}*)$\sim$14.5.  Assuming that
[\ion{Si}{ii}/H]=[\ion{S}{ii}/H], i.e. Si is undepleted onto dust
grains, we obtain log N(\ion{Si}{ii})$\sim$16.2 (as log
N(\ion{S}{ii})=15.84 and the Solar abundance difference between S and
Si is 0.34) and n$_{3/2}$/n$_{1/2}\sim$ --1.7. This ratio corresponds
to a volume density of n$_{\ion{H}{i}} \sim$ 100 cm$^{-3}$ when the
free electron density (n$_{\rm e}$) is assumed to be 10\% of that of
the \ion{H}{i} density (n$_{\ion{H}{i}}$), and n$_{\ion{H}{i}} \sim
10^4$ cm$^{-3}$ when n$_{\rm e} \leq 10^{-4}$ n$_{\ion{H}{i}}$
\citep[see Fig.  8 of][]{2002MNRAS.329..135S}. If these fine-structure
lines originate in the same region as the neutral hydrogen, then the
\ion{Si}{ii}* medium is mostly neutral and the free electrons will
mainly come from ionization of neutral atoms with an ionization
potential lower than 13.6 eV, whose solar abundance relative to
hydrogen is typically 10$^{-4}$ \citep[see][]{2002MNRAS.329..135S}.
This would result in an expected ratio n$_{\rm e} \sim 10^{-4}$
n$_{\ion{H}{i}}$.  We have assumed that the fine-structure levels are
populated by collisions between particles, and not through direct
excitation by infra-red photons (although this mechanism is probably
not important in the case of \ion{Si}{ii}*), or fluorescence
\citep{2002MNRAS.329..135S}.  Under this assumption, we can divide the
column density by the volume density to obtain an order of magnitude
estimate of the size (diameter) of the absorbing region: $\sim$ 5pc
(taking n$_{\ion{H}{i}}=10^3$ cm$^{-3}$). As a comparison, Galactic
molecular cloud sizes range from roughly 0.5pc to 50pc
\citep{1987ApJ...319..730S}. Following \citet{2002MNRAS.329..135S}, we
can also estimate the mass of the \ion{Si}{ii}* absorbing cloud, $\rm
M=m_{\rm p}N(\ion{H}{i})l_{\ion{Si}{ii}*}^2$ (where $\rm m_{\rm p}$ is
the proton mass and $\rm l_{\ion{Si}{ii}*}$ is the diameter of the
\ion{Si}{ii}* absorbing region) to be $\rm M=2\times 10^3$ M$\subsun$.
However, the size and mass estimates would be upper limits if the
\ion{Si}{ii}* ions are only partly associated with the entire
\ion{H}{i} column and/or \ion{Si}{ii} column. For instance, if half of
the \ion{Si}{ii} absorption would not be related to the \ion{Si}{ii}*
absorbing region, the actual volume density would be roughly twice as
large, the corresponding cloud size twice as small and the mass a
factor of four smaller. If, on the other hand, fluorescence plays a
non-negligible role, the size and mass estimates above would be lower
limits.

\begin{figure*}[tp]
  \centering
  \includegraphics[width=9cm]{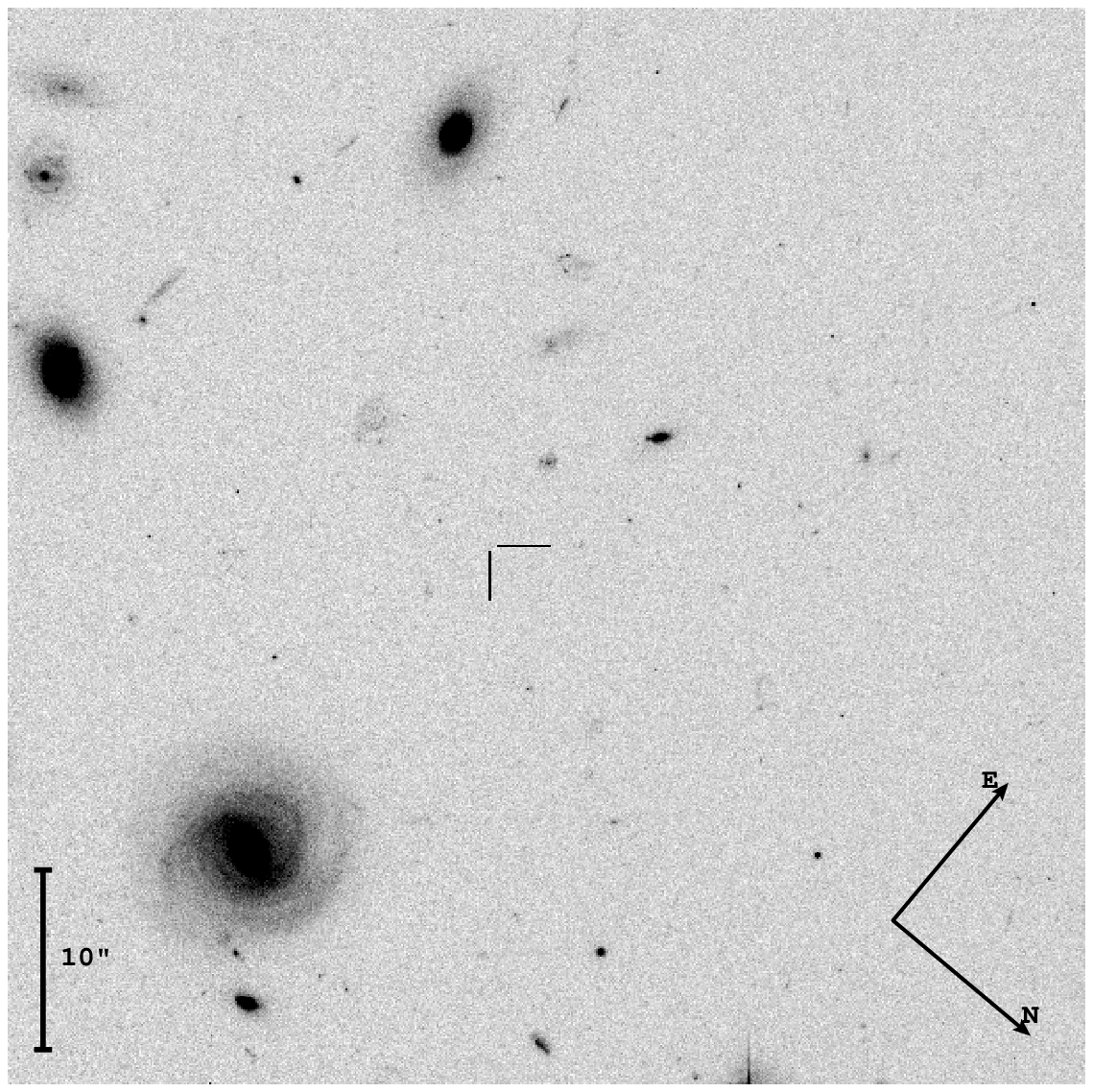}\includegraphics[width=9cm]{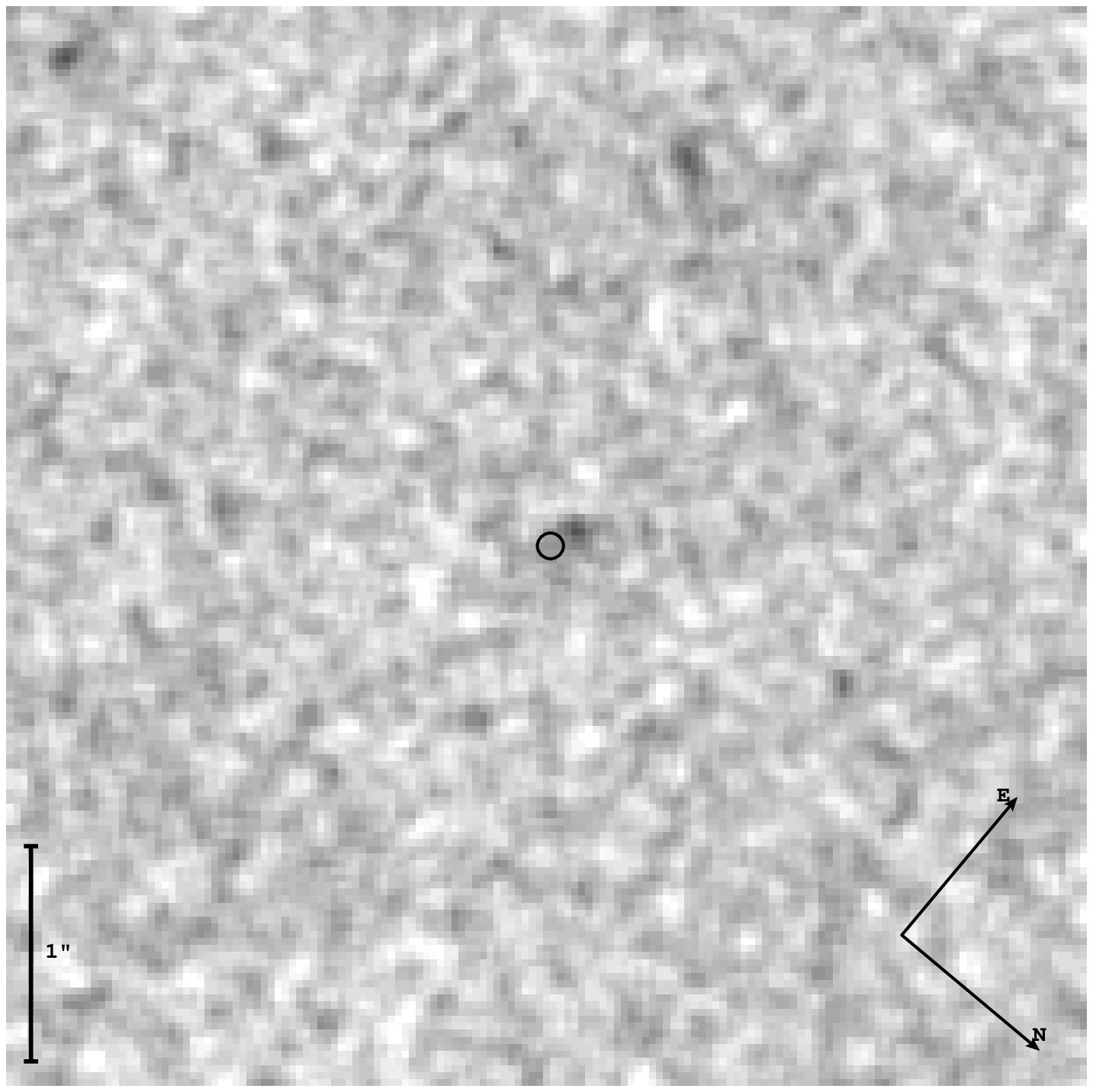}
  \caption{HST/ACS F606W imaging of the field of \grb. On
    the left panel the 1\arcmin$\times$1\arcmin\ field surrounding the
    burst is shown, whose location is indicated by the cross hair.  On
    the right panel the central 5\arcsec$\times$5\arcsec\ region is
    shown, where in this case the image has been median filtered using
    a box with 3 pixels on a side. The position of the early afterglow
    and its uncertainty (0\farcs06 in radius) is depicted by the black
    circle. At a distance of 0\farcs14 (1kpc), an extended object is
    detected. This is the probable host galaxy of \grb\, for which we
    measure AB(F606W)=28.0$\pm$0.3 mag.
      \label{hst}}
\end{figure*}

\section{Ly$\alpha$ in emission}
\label{sec:lya}

As can be seen in Fig. \ref{spectrum}, in the \lya\ trough \lya\ in
emission is detected. Using {\it splot} within IRAF, we measured the
center of the line at 5316.9$\pm$0.7\AA, corresponding to
$z$=3.3736$\pm$0.0006, with a FWHM of 237 km s$^{-1}$. The velocity
offset with respect to the metal-absorption lines is +151$\pm$46 km
s$^{-1}$, i.e.  the \lya\ emitting region is red-shifted with respect
to the material responsible for the absorption lines. We measured a
\lya\ flux of F = (1.2$\pm$0.1)$\times$10$^{-17}$ erg cm$^{-2}$
s$^{-1}$ in the 1400V spectrum. The emission line is also detected,
albeit barely, in the lower resolution 300V spectrum with F =
(1.0$\pm$0.3)$\times$10$^{-17}$ erg cm$^{-2}$ s$^{-1}$. Assuming $\rm
H_0 = 70 \,km\,s^{-1}\,Mpc^{-1}$, $\Omega_{\rm M}$ = 0.3 and
$\Omega_{\Lambda}$ = 0.7, $z$=3.372 corresponds to a luminosity
distance of d$_L$=9$\times$10$^{28}$ cm.  This transforms the observed
flux into a \lya\ luminosity of L$_{\rm Ly\alpha}$ =
(1.2$\pm$0.1)$\times$10$^{42}$ erg s$^{-1}$.

Adopting a relation between measured \lya\ luminosity and
star-formation rate of L/SFR = 10$^{42}$ erg s$^{-1}$ per 1 M\subsun\ 
\citep{1998ARA&A..36..189K,1998AJ....115.1319C}, the \lya\ 
star-formation rate in the \grb\ host galaxy is roughly 1.2 M\subsun\ 
yr$^{-1}$.  This value has not been corrected for extinction and hence
is a lower limit.  The \lya\ luminosity is roughly twice as large as
the one obtained for the host of GRB\,971214 at $z$=3.42:
(0.66$\pm$0.07)$\times$10$^{42}$ erg s$^{-1}$
\citep{1998Natur.393...35K} but is at the low end of a sample of 10
\lya\ emitters found at $z\sim3.4$ by \citet{1998AJ....115.1319C} in
and around the HDF and Hawaii deep field SSA 22 (with L$_{\rm
  Ly\alpha}$ ranging from 1.2 to 8$\times$10$^{42}$ erg s$^{-1}$).
However, a deeper survey by \citet{fynbobridge} has resulted in 42
confirmed \lya\ emitting galaxies at $z\sim3$, with most luminosities
ranging from 0.4-2$\times$10$^{42}$ erg s$^{-1}$.

\lya\ emission has also been observed in the troughs of half a dozen
of QSO-DLAs
\citep[e.g.][]{1993A&A...270...43M,1995qal..conf...55P,1996Natur.382..234D,1998A&A...330...19M,1999MNRAS.305..849F,2002ApJ...574...51M}.
It is believed that this emission originates in the DLA host galaxy
itself, and not in QSO photo-ionized regions when $z_{\rm abs}\sim
z_{\rm QSO}$. In the \grb\ case, it is clear that the emission is
produced by photo-ionization by massive stars (not necessarily related
to the GRB) in the host galaxy.  An origin in the immediate
environment of the GRB is not possible, as this emission would be
absorbed equally well as the afterglow continuum emission around \lya\ 
by the high neutral hydrogen column density along this sightline.

\section{HST imaging of the host galaxy}
\label{sec:hst}

The field of GRB 030323 was observed for 4$\times$480s with HST/ACS in
the F606W filter on July 20, 2003, starting at 23:00 UT. The dithered
exposures were drizzled with the multidrizzle routine\footnote{see
  http://stsdas.stsci.edu/pydrizzle/multidrizzle} to produce an output
image with a scale of 0\farcs033 per pixel. Fig. \ref{hst} shows the
1\arcmin$\times$1\arcmin\ field of \grb\ on the left panel, and on the
right panel the central 5\arcsec$\times$5\arcsec\ region. The close-up
image has been convolved with a median filter of 3 by 3 pixels,
excluding the central pixel of the kernel in the median calculation.
The position of the early afterglow has been projected onto the HST
image using the 3min FORS2 V-band image of March 26.35 (see Table
\ref{tab:imaging}); 8 objects were used to perform the transformation,
with a resulting positional accuracy of 0\farcs06. This error circle
is shown on the zoomed image.

At 4.3 pixels, or 0\farcs14, an extended object is detected, which can
be identified as the probable host of \grb.  The Gaussian FWHM of the
point spread function of this object (5.0 pixels) is significantly
greater than that of stars in the field (2.8 pixels), and SExtractor
\citep{1996A&AS..117..393B} classifies it as a galaxy, with the
star-galaxy classification flag equal to 0.03 (which normally is 0 for
a definite galaxy and 1 for a definite star). Adopting H$_0$=70 km
s$^{-1}$ Mpc$^{-1}$, $\Omega_{\rm M}$=0.3 and $\Omega_{\Lambda}$=0.7,
0\farcs14 at $z=3.372$ corresponds to an angular diameter distance of
1 kpc \citep[see][]{hogg}.

For the host galaxy, we measured a magnitude AB(F606W)=28.15$\pm$0.11
mag using SExtractor's isophotal flux estimate
\citep{1996A&AS..117..393B}, while we obtain AB(F606W)=28.1$\pm$0.2
mag with aperture photometry using an aperture radius of 5 pixels.
With a 10-pixel aperture radius, the magnitude increases to
AB(F606W)=27.8 mag.  We adopted a zeropoint of 26.51 mag for the
conversion of counts to AB magnitudes. The error estimates given above
only include the Poisson errors. Assuming that the host galaxy has a
flat spectrum in AB (i.e.  $\beta$=0 in F$_{\nu}\propto\nu^{\beta}$),
the AB magnitude that we finally adopt: AB(F606W)=28.0$\pm$0.3 mag, is
the same when converting it to the Johnson V band.

From the UV continuum emission of the host galaxy we can obtain
another crude estimate of the star-formation rate, independent of the
SFR inferred from \lya\ in emission (see Sect. \ref{sec:lya}).  The
magnitude AB(F606W)=28 mag corresponds to a flux F$_{\nu}$=2.3$\times
10^{-31}$ erg s$^{-1}$ cm$^{-2}$ Hz$^{-1}$, which at the luminosity
distance of \grb, d$_L$=9$\times$10$^{28}$ cm, results in
L$_{\nu}$=5.4$\times 10^{27}$ erg s$^{-1}$ Hz$^{-1}$.  Using the
SFR-L$_{\nu}$(1500-2800\AA) relation of \citet{1998ARA&A..36..189K},
this transforms to a SFR(UV) of 0.7 M\subsun\ yr$^{-1}$. We note that
we have neglected \lya\ forest absorption at the blue end of the F606W
filter (4700-7200\AA, while \lya\ is at 5317\AA), which would increase
the SFR estimate by roughly 25\% to around 0.9 M\subsun\ yr$^{-1}$,
which is very similar to the estimate from the \lya\ emission line
(1.2 M\subsun\ yr$^{-1}$).

\section{Discussion}
\label{sec:discussion}

Our observations show that \grb\ occurred behind a very high
\ion{H}{i} column density, in an environment (immediate and
host-galaxy combined) having a low molecular hydrogen fraction
(f$\lsim$10$^{-6}$), a low metallicity ([S/H]=--1.26$\pm$0.20) and a
low dust content ($\kappa$=0.02). For the DLA host of GRB\,020124,
\citet{hjorth020124} also find evidence for a large \ion{H}{i} column
density with a low reddening.  The inferred low dust content may be
interpreted as a selection bias: GRBs that would occur in a dusty host
galaxy would be harder to detect because they would then be fainter.
However, in this case one would expect to observe many GRB afterglows
with considerable extinction in the optical, for which there is no
clear evidence \citep{2001ApJ...549L.209G,2002MNRAS.330..583L}. In
apparent contradiction with this is the detection of several host
galaxies in the radio and sub-mm
\citep[e.g.][]{frail222,2003MNRAS.338....1B,2003ApJ...588...99B},
suggesting that at least some GRB hosts are dusty, as expected when
most of the star formation in the universe occurs in submm-bright
galaxies \citep[see][]{2002MNRAS.329..465R}.  Dust destruction
\citep[e.g.][]{2001ApJ...563..597F,2002ApJ...569..780D,2003ApJ...585..775P},
which has been proposed to account for the apparent discrepancy
between the low optical extinction and large (X-ray) gas column
densities \citep{2001ApJ...549L.209G}, could play a role, but is not
required by our data: the reduced metallicity and hence the low
dust-to-gas ratio in the host of \grb\ is sufficient to explain the
combination of a large \ion{H}{i} column density with a low optical
extinction \citep[see also][]{hjorth020124}.

\lya\ in emission is detected for \grb, and we inferred a
star-formation rate of about 1 M\subsun\ yr$^{-1}$, which is in good
agreement with the SFR value that we obtained from the UV continuum
emission of the host galaxy. \citet{fynbolya} note that \lya\ is
commonly observed in all GRB host galaxies at high redshift for which
it could be detected.  In contrast, only 25\% of the Lyman-break
galaxies are \lya\ emitters with an equivalent width EW $>$ 20 \AA\ 
\citep{2003ApJ...588...65S}.  \citet{fynbolya} suggest that this
difference is due to GRB hosts having a low metallicity and a low dust
content, consistent with our observations of \grb\ and with those of
GRB\,020124 \citep{hjorth020124}.  We note that QSO-DLAs also have a
low metallicity and a low dust content, but they rarely show \lya\ in
emission. However, since most galaxy counterparts of QSO-DLAs are very
faint, \lya\ in emission is not expected to be detected in most cases
with the current detection limits \citep[see][]{1999MNRAS.305..849F}.
The low-dust inference for GRB\,020124 \citep{hjorth020124} and \grb\ 
is different from the results of \citet{2003ApJ...585..638S}, who find
evidence for a high dust content in three GRB host galaxies.

From the fine-structure lines \ion{Si}{ii}* \lam\lam 1309,1533 , we
estimated the \ion{H}{i} volume density of the gas producing this
absorption: n$_{\ion{H}{i}}=10^2-10^4$ cm$^{-3}$, under the assumption
that these fine-structure levels are populated by collisions, and not
through direct excitation by infra-red photons (which is not an
important excitation mechanism in the case of \ion{Si}{ii}*), or
fluorescence \citep[see][]{2002MNRAS.329..135S}.  This volume density
is higher than inferred for QSO-DLA environments
\citep{2002MNRAS.329..135S}, but typical of Galactic molecular clouds
\citep[e.g.][]{1999osps.conf....3B,2002ApJ...565..174R}. As these
lines has never been clearly detected up to now in QSO-DLAs, the
detection of these {Si}{\sc ii}* lines in the \grb\ spectrum suggests
an origin in the vicinity of the GRB place of birth (e.g. the
star-forming region in which it exploded). Combining the measured
\ion{H}{i} column density with the order of magnitude estimate of the
\ion{H}{i} volume density, we obtain a size (diameter) of $\sim$ 5pc
(taking n$_{\ion{H}{i}}=10^3$ cm$^{-3}$) and a mass of $\sim$
2$\times$10$^3$ M\subsun\ for the \ion{Si}{ii}* absorbing region.

With the volume density so high, one would expect hydrogen molecules
to be present, which, surprisingly, we do not detect.  We obtain a
rather strong upper limit on the mean molecular fraction of the gas in
the GRB environment and the host galaxy:
f$\equiv$2$N$(H$_2$)/(2$N$(H$_2$)+$N$(\ion{H}{i}))$\lsim$10$^{-6}$.
This could be explained by the low metallicity of the gas
\citep[see][]{ledoux}, but it may also be that the molecules in the
GRB environment have been dissociated by the strong GRB UV/X-ray
emission \citep[e.g.][]{2002ApJ...569..780D}. In the latter case,
however, the UV/X-ray flash would also ionize the neutral gas in the
GRB vicinity \citep[see][]{2002ApJ...569..780D}, which would make the
high \ion{H}{i} column density detection improbable. Therefore, a
large fraction of the \ion{H}{i} column density may not be located
close to the GRB explosion site, but elsewhere in the host galaxy,
while the high volume density \ion{Si}{ii}* region (and the expected
molecular hydrogen), is located in the vicinity of the burst. In this
case, the disks of GRB host galaxies need to be much denser than the
Galactic disk, as 7 random sight lines through the disk toward the
location of the Earth would not result in 5 \ion{H}{i} column
densities above 10$^{21}$ cm$^{-2}$ \citep[see Fig.  5
of][]{1990ARA&A..28..215D}, as is observed for GRB sightlines (see
Fig. \ref{histo}). Finally, the population of the \ion{Si}{ii}* levels
may have been partly caused by fluorescence of photons from the GRB
itself, in which case the volume density estimate above is a strict
upper limit.

The \ion{H}{i} column density that we inferred toward \grb\ is higher
than that of any (QSO- or GRB-) DLA measured using \lya\ in
absorption. It is generally assumed that the apparent \ion{H}{i}
column density limit of N(\ion{H}{i})$\sim$10$^{22}$ atoms cm$^{-2}$
for QSO-DLAs is due to an observational bias against the detection of
such high-column density systems, as these would obscure the
background QSO if they contain some dust
\citep[e.g.][]{1984ApJ...278....1O,1993ApJ...402..479F}.  However, a
radio-selected QSO survey for DLA systems by
\citet{2001A&A...379..393E} did not uncover a previously unrecognized
population of N(\ion{H}{i})$>10^{21}$ cm$^{-2}$ DLA systems in front
of faint QSOs. An alternative scenario was proposed by
\citet{2001ApJ...562L..95S}: the lack of high \ion{H}{i} column
density systems could be due to the conversion of \ion{H}{i} to H$_2$
as the neutral gas density increases. This picture is consistent with
observations of Galactic molecular clouds
\citep[e.g.][]{1999osps.conf....3B}. In \grb, however, we do not find
any evidence for the presence of H$_2$ in addition to \ion{H}{i} to
support this scenario.  Future GRBs with possibly even larger
\ion{H}{i} column densities than that toward \grb\ could provide
further constraints to the existence of a rapid conversion of
\ion{H}{i} to H$_2$ at high \ion{H}{i} column densities.

We compared the metallicities and \ion{H}{i} column densities of the
(still very small) sample of GRB-DLAs with QSO-DLAs, and we found both
quantities to be higher in GRB-DLAs than in QSO-DLAs.  This is not
surprising, as GRBs are now known to probe massive-star forming
regions \citep{stanek,hjorth030329} where the gas density and the
metallicity are higher than along random QSO sight lines through
foreground galaxies.  A KS test applied to the column densities shows
that the probability that the GRB- and QSO-DLA samples are drawn from
the same parent distribution is very low (0.0006). On the other hand,
two GRB afterglows have very low column densities.  A large sample of
high-resolution spectra of GRB afterglows could provide statistical
information about the distribution of the gas in high-redshift
star-forming regions, in addition to the evolution of the metallicity
and dust and H$_2$ contents of GRB host galaxies.  Such a sample can
be created in the years to come thanks to rapid and accurate GRB
localizations from future satellite missions such as
Swift\footnote{see http://swift.gsfc.nasa.gov/} and EXIST\footnote{see
  http://exist.gsfc.nasa.gov/}.

\begin{acknowledgements}
  We thank Sylvio Klose, Jochen Greiner and Martin Zwaan for helpful
  comments, and the referee, Sandra Savaglio, for an excellent report.
  We acknowledge benefits from collaboration within the Research
  Training Network "Gamma-Ray Bursts: An Enigma and a Tool", funded by
  the EU under contract HPRN-CT-2002-00294.  ER acknowledges support
  from NWO grant nr. 614-51-003. JPUF gratefully acknowledges support
  from the Carlsberg Foundation. This work was supported by the Danish
  Natural Science Research Council (SNF).  JMCC acknowledges the
  receipt of a FPI doctoral fellowship from Spain's Ministerio de
  Ciencia y Tecnolog\'{\i}a.  This publication makes use of data
  products from the Two Micron All Sky Survey, which is a joint
  project of the University of Massachusetts and the Infrared
  Processing and Analysis Center/California Institute of Technology,
  funded by the National Aeronautics and Space Administration and the
  National Science Foundation.
\end{acknowledgements}

\bibliographystyle{aa} 
\bibliography{references}

\begin{thebibliography}{102}
\expandafter\ifx\csname natexlab\endcsname\relax\def\natexlab#1{#1}\fi

\bibitem[{{Akerlof} {et~al.}(1999){Akerlof}, {Balsano}, {Barthelemy}, {Bloch},
  {Butterworth}, {Casperson}, {Cline}, {Fletcher}, {Frontera}, {Gisler},
  {Heise}, {Hills}, {Kehoe}, {Lee}, {Marshall}, {McKay}, {Miller}, {Piro},
  {Priedhorsky}, {Szymanski}, \& {Wren}}]{1999Natur.398..400A}
{Akerlof}, C., {Balsano}, R., {Barthelemy}, S., {et~al.} 1999, \nat, 398, 400

\bibitem[{{Andersen} {et~al.}(1999){Andersen}, {Castro-Tirado}, {Hjorth},
  {M{\o}ller}, {Pedersen}, {Caon}, {Marina Cairos}, {Korhonen}, {Zapatero
  Osorio}, {Perez}, \& {Frontera}}]{1999Sci...283.2075A}
{Andersen}, M.~I., {Castro-Tirado}, A.~J., {Hjorth}, J., {et~al.} 1999,
  Science, 283, 2075

\bibitem[{{Andersen} {et~al.}(2000){Andersen}, {Hjorth}, {Pedersen}, {Jensen},
  {Hunt}, {Gorosabel}, {M{\o}ller}, {Fynbo}, {Kippen}, {Thomsen}, {Olsen},
  {Christensen}, {Vestergaard}, {Masetti}, {Palazzi}, {Hurley}, {Cline},
  {Kaper}, \& {Jaunsen}}]{2000A&A...364L..54A}
{Andersen}, M.~I., {Hjorth}, J., {Pedersen}, H., {et~al.} 2000, \aap, 364, L54

\bibitem[{{Bahcall} \& {Wolf}(1968)}]{1968ApJ...152..701B}
{Bahcall}, J.~N. \& {Wolf}, R.~A. 1968, \apj, 152, 701

\bibitem[{{Barnard} {et~al.}(2003){Barnard}, {Blain}, {Tanvir}, {Natarajan},
  {Smith}, {Wijers}, {Kouveliotou}, {Rol}, {Tilanus}, \&
  {Vreeswijk}}]{2003MNRAS.338....1B}
{Barnard}, V.~E., {Blain}, A.~W., {Tanvir}, N.~R., {et~al.} 2003, \mnras, 338,
  1

\bibitem[{{Berger} {et~al.}(2003){Berger}, {Cowie}, {Kulkarni}, {Frail},
  {Aussel}, \& {Barger}}]{2003ApJ...588...99B}
{Berger}, E., {Cowie}, L.~L., {Kulkarni}, S.~R., {et~al.} 2003, \apj, 588, 99

\bibitem[{{Bertin} \& {Arnouts}(1996)}]{1996A&AS..117..393B}
{Bertin}, E. \& {Arnouts}, S. 1996, \aaps, 117, 393

\bibitem[{{Blitz} \& {Williams}(1999)}]{1999osps.conf....3B}
{Blitz}, L. \& {Williams}, J.~P. 1999, in NATO ASIC Proc. 540: The Origin of
  Stars and Planetary Systems, 3

\bibitem[{{Boiss\'e} \& {Bergeron}(1985)}]{1985A&A...145...59B}
{Boiss\'e}, P. \& {Bergeron}, J. 1985, \aap, 145, 59

\bibitem[{{Bouchet} {et~al.}(1985){Bouchet}, {Lequeux}, {Maurice}, {Pr\'evot},
  \& {Pr\'evot-Burnichon}}]{1985A&A...149..330B}
{Bouchet}, P., {Lequeux}, J., {Maurice}, E., {Pr\'evot}, L., \&
  {Pr\'evot-Burnichon}, M.~L. 1985, \aap, 149, 330

\bibitem[{{Castro} {et~al.}(2003){Castro}, {Galama}, {Harrison}, {Holtzman},
  {Bloom}, {Djorgovski}, \& {Kulkarni}}]{castro926}
{Castro}, S., {Galama}, T.~J., {Harrison}, F.~A., {et~al.} 2003, \apj, 586, 128

\bibitem[{{Cohen} {et~al.}(2003){Cohen}, {Wheaton}, \&
  {Megeath}}]{2003AJ....126.1090C}
{Cohen}, M., {Wheaton}, W.~A., \& {Megeath}, S.~T. 2003, \aj, 126, 1090

\bibitem[{{Cowie} \& {Hu}(1998)}]{1998AJ....115.1319C}
{Cowie}, L.~L. \& {Hu}, E.~M. 1998, \aj, 115, 1319

\bibitem[{{Cristiani} {et~al.}(1993){Cristiani}, {Giallongo}, {Buson},
  {Gouiffes}, \& {La Franca}}]{1993A&A...268...86C}
{Cristiani}, S., {Giallongo}, E., {Buson}, L.~M., {Gouiffes}, C., \& {La
  Franca}, F. 1993, \aap, 268, 86

\bibitem[{{Curran} {et~al.}(2002){Curran}, {Webb}, {Murphy}, {Bandiera},
  {Corbelli}, \& {Flambaum}}]{2002PASA...19..455C}
{Curran}, S.~J., {Webb}, J.~K., {Murphy}, M.~T., {et~al.} 2002, Publications of
  the Astronomical Society of Australia, 19, 455

\bibitem[{{Dado} {et~al.}(2002){Dado}, {Dar}, \& {De R{\'
  u}jula}}]{2002A&A...388.1079D}
{Dado}, S., {Dar}, A., \& {De R{\' u}jula}, A. 2002, \aap, 388, 1079

\bibitem[{{Dado} {et~al.}(2003{\natexlab{a}}){Dado}, {Dar}, \& {De R{\'
  u}jula}}]{2003ApJ...585L..15D}
{Dado}, S., {Dar}, A., \& {De R{\' u}jula}, A. 2003{\natexlab{a}}, \apjl, 585,
  L15

\bibitem[{{Dado} {et~al.}(2003{\natexlab{b}}){Dado}, {Dar}, \& {De R{\'
  u}jula}}]{2003ApJ...594L..89D}
{Dado}, S., {Dar}, A., \& {De R{\' u}jula}, A. 2003{\natexlab{b}}, \apjl, 594,
  L89

\bibitem[{{Dai} \& {Cheng}(2001)}]{2001ApJ...558L.109D}
{Dai}, Z.~G. \& {Cheng}, K.~S. 2001, \apjl, 558, L109

\bibitem[{{Dickey} \& {Lockman}(1990)}]{1990ARA&A..28..215D}
{Dickey}, J.~M. \& {Lockman}, F.~J. 1990, \araa, 28, 215

\bibitem[{{Diplas} \& {Savage}(1994)}]{1994ApJ...427..274D}
{Diplas}, A. \& {Savage}, B.~D. 1994, \apj, 427, 274

\bibitem[{{Djorgovski} {et~al.}(1996){Djorgovski}, {Pahre}, {Bechtold}, \&
  {Elston}}]{1996Natur.382..234D}
{Djorgovski}, S.~G., {Pahre}, M.~A., {Bechtold}, J., \& {Elston}, R. 1996,
  \nat, 382, 234

\bibitem[{{Draine} \& {Hao}(2002)}]{2002ApJ...569..780D}
{Draine}, B.~T. \& {Hao}, L. 2002, \apj, 569, 780

\bibitem[{{Ellison} {et~al.}(2001){Ellison}, {Yan}, {Hook}, {Pettini}, {Wall},
  \& {Shaver}}]{2001A&A...379..393E}
{Ellison}, S.~L., {Yan}, L., {Hook}, I.~M., {et~al.} 2001, \aap, 379, 393

\bibitem[{{Fall} \& {Pei}(1993)}]{1993ApJ...402..479F}
{Fall}, S.~M. \& {Pei}, Y.~C. 1993, \apj, 402, 479

\bibitem[{{Frail} {et~al.}(2002){Frail}, {Bertoldi}, {Moriarty-Schieven},
  {Berger}, {Price}, {Bloom}, {Sari}, {Kulkarni}, {Gerardy}, {Reichart},
  {Djorgovski}, {Galama}, {Harrison}, {Walter}, {Shepherd}, {Halpern}, {Peck},
  {Menten}, {Yost}, \& {Fox}}]{frail222}
{Frail}, D.~A., {Bertoldi}, F., {Moriarty-Schieven}, G.~H., {et~al.} 2002,
  \apj, 565, 829

\bibitem[{{Fruchter} {et~al.}(2001){Fruchter}, {Krolik}, \&
  {Rhoads}}]{2001ApJ...563..597F}
{Fruchter}, A., {Krolik}, J.~H., \& {Rhoads}, J.~E. 2001, \apj, 563, 597

\bibitem[{{Fynbo} {et~al.}(2001){Fynbo}, {Gorosabel}, {M{\o}ller}, {Hjorth},
  {Andersen}, {Egholm}, {Jensen}, {Pedersen}, {Thomsen}, \&
  {Weidinger}}]{fynbo926spectrum}
{Fynbo}, J.~P.~U., {Gorosabel}, J., {M{\o}ller}, P., {et~al.} 2001, in
  Lighthouses of the Universe: The Most Luminous Celestial Objects and Their
  Use for Cosmology, Proc. of the MPA/ESO/MPE/USM Joint Astronomy Conf., eds.
  M. Gilfanov, R. Sunyaev, \& E. Churazov (Garching: Springer), p.187; preprint
  astro-ph/0110603

\bibitem[{{Fynbo} {et~al.}(2003{\natexlab{a}}){Fynbo}, {Jakobsson},
  {M{\o}ller}, {Hjorth}, {Thomsen}, {Andersen}, {Fruchter}, {Gorosabel},
  {Holland}, {Ledoux}, {Pedersen}, {Rhoads}, {Weidinger}, \&
  {Wijers}}]{fynbolya}
{Fynbo}, J.~P.~U., {Jakobsson}, P., {M{\o}ller}, P., {et~al.}
  2003{\natexlab{a}}, \aap, 406, L63

\bibitem[{{Fynbo} {et~al.}(2003{\natexlab{b}}){Fynbo}, {Ledoux}, {M{\o}ller},
  {Thomsen}, \& {Burud}}]{fynbobridge}
{Fynbo}, J.~P.~U., {Ledoux}, C., {M{\o}ller}, P., {Thomsen}, B., \& {Burud}, I.
  2003{\natexlab{b}}, \aap, 407, 147

\bibitem[{{Fynbo} {et~al.}(1999){Fynbo}, {M{\o}ller}, \&
  {Warren}}]{1999MNRAS.305..849F}
{Fynbo}, J.~P.~U., {M{\o}ller}, P., \& {Warren}, S.~J. 1999, \mnras, 305, 849

\bibitem[{{Galama} {et~al.}(1998{\natexlab{a}}){Galama}, {Groot}, {Van
  Paradijs}, {Kouveliotou}, {Strom}, {Wijers}, {Tanvir}, {Bloom}, {Centurion},
  {Telting}, {Rutten}, {Smith}, {Mackey}, {Smartt}, {Benn}, {Heise}, \& {in 't
  Zand}}]{1998ApJ...497L..13G}
{Galama}, T.~J., {Groot}, P.~J., {Van Paradijs}, J., {et~al.}
  1998{\natexlab{a}}, \apjl, 497, L13

\bibitem[{{Galama} {et~al.}(1998{\natexlab{b}}){Galama}, {Vreeswijk}, {Van
  Paradijs}, {Kouveliotou}, {Augusteijn}, {Bohnhardt}, {Brewer}, {Doublier},
  {Gonzalez}, {Leibundgut}, {Lidman}, {Hainaut}, {Patat}, {Heise}, {in 't
  Zand}, {Hurley}, {Groot}, {Strom}, {Mazzali}, {Iwamoto}, {Nomoto}, {Umeda},
  {Nakamura}, {Young}, {Suzuki}, {Shigeyama}, {Koshut}, {Kippen}, {Robinson},
  {de Wildt}, {Wijers}, {Tanvir}, {Greiner}, {Pian}, {Palazzi}, {Frontera},
  {Masetti}, {Nicastro}, {Feroci}, {Costa}, {Piro}, {Peterson}, {Tinney},
  {Boyle}, {Cannon}, {Stathakis}, {Sadler}, {Begam}, \&
  {Ianna}}]{1998Natur.395..670G}
{Galama}, T.~J., {Vreeswijk}, P.~M., {Van Paradijs}, J., {et~al.}
  1998{\natexlab{b}}, \nat, 395, 670

\bibitem[{{Galama} \& {Wijers}(2001)}]{2001ApJ...549L.209G}
{Galama}, T.~J. \& {Wijers}, R. A. M.~J. 2001, \apjl, 549, L209

\bibitem[{{Garnavich} {et~al.}(2000){Garnavich}, {Loeb}, \&
  {Stanek}}]{2000ApJ...544L..11G}
{Garnavich}, P.~M., {Loeb}, A., \& {Stanek}, K.~Z. 2000, \apjl, 544, L11

\bibitem[{{Gilmore} {et~al.}(2003){Gilmore}, {Kilmartin}, \&
  {Henden}}]{2003GCN..1949....1G}
{Gilmore}, A., {Kilmartin}, P., \& {Henden}, A. 2003, GRB Circular Network,
  1949

\bibitem[{{Granot} {et~al.}(2003){Granot}, {Nakar}, \& {Piran}}]{granot030329}
{Granot}, J., {Nakar}, E., \& {Piran}, T. 2003, \nat, 426, 138

\bibitem[{{Graziani} {et~al.}(2003){Graziani}, {Shirasaki}, {Matsuoka},
  {Tamagawa}, {Torii}, {Sakamoto}, {Suzuki}, {Yoshida}, {Fenimore}, {Galassi},
  {Tavenner}, {Donaghy}, {Nakagawa}, {Takahashi}, {Satoh}, {Urata}, {Ricker},
  {Atteia}, {Kawai}, {Lamb}, {Woosley}, {Vanderspek}, {Villasenor}, {Crew},
  {Doty}, {Monnelly}, {Butler}, {Cline}, {Jernigan}, {Levine}, {Martel},
  {Morgan}, {Prigozhin}, {Azzibrouck}, {Braga}, {Manchanda}, {Pizzichini},
  {Boer}, {Olive}, {Dezalay}, {Barraud}, \& {Hurley}}]{2003GCN..1956....1G}
{Graziani}, C., {Shirasaki}, Y., {Matsuoka}, M., {et~al.} 2003, GRB Circular
  Network, 1956

\bibitem[{{Greiner} {et~al.}(2003){Greiner}, {Peimbert}, {Estaban}, {Kaufer},
  {Jaunsen}, {Smoke}, {Klose}, \& {Reimer}}]{2003GCN..2020....1G}
{Greiner}, J., {Peimbert}, M., {Estaban}, C., {et~al.} 2003, GRB Circular
  Network, 2020

\bibitem[{{Grevesse} \& {Sauval}(1998)}]{1998SSRv...85..161G}
{Grevesse}, N. \& {Sauval}, A.~J. 1998, Space Science Reviews, 85, 161

\bibitem[{{Haehnelt} {et~al.}(1998){Haehnelt}, {Steinmetz}, \&
  {Rauch}}]{1998ApJ...495..647H}
{Haehnelt}, M.~G., {Steinmetz}, M., \& {Rauch}, M. 1998, \apj, 495, 647

\bibitem[{{Henden}(2003)}]{2003GCN..1948....1H}
{Henden}, A. 2003, GRB Circular Network, 1948

\bibitem[{{Hjorth} {et~al.}(2003{\natexlab{a}}){Hjorth}, {M{\o}ller},
  {Gorosabel}, {Fynbo}, {Toft}, {Jaunsen}, {Kaas}, {Pursimo}, {Torii}, {Kato},
  {Yamaoka}, {Yoshida}, {Thomsen}, {Andersen}, {Burud}, {Cer{\' o}n},
  {Castro-Tirado}, {Fruchter}, {Kaper}, {Kouveliotou}, {Masetti}, {Palazzi},
  {Pedersen}, {Pian}, {Rhoads}, {Rol}, {Tanvir}, {Vreeswijk}, {Wijers}, \& {van
  den Heuvel}}]{hjorth020124}
{Hjorth}, J., {M{\o}ller}, P., {Gorosabel}, J., {et~al.} 2003{\natexlab{a}},
  \apj, 597, 699

\bibitem[{{Hjorth} {et~al.}(2003{\natexlab{b}}){Hjorth}, {Sollerman},
  {M{\o}ller}, {Fynbo}, {Woosley}, {Kouveliotou}, {Tanvir}, {Greiner}, \&
  {Andersen}}]{hjorth030329}
{Hjorth}, J., {Sollerman}, J., {M{\o}ller}, P., {et~al.} 2003{\natexlab{b}},
  \nat, 423, 847

\bibitem[{{Hogg}(1999)}]{hogg}
{Hogg}, D.~W. 1999, astro-ph/9905116

\bibitem[{{Holland} {et~al.}(2003){Holland}, {Weidinger}, {Fynbo}, {Gorosabel},
  {Hjorth}, {Pedersen}, {Alvarez}, {Augusteijn}, {Cer{\' o}n}, {Castro-Tirado},
  {Dahle}, {Egholm}, {Jakobsson}, {Jensen}, {Levan}, {M{\o}ller}, {Pedersen},
  {Pursimo}, {Ruiz-Lapuente}, \& {Thomsen}}]{2003AJ....125.2291H}
{Holland}, S.~T., {Weidinger}, M., {Fynbo}, J.~P.~U., {et~al.} 2003, \aj, 125,
  2291

\bibitem[{{Jensen} {et~al.}(2001){Jensen}, {Fynbo}, {Gorosabel}, {Hjorth},
  {Holland}, {M{\o}ller}, {Thomsen}, {Bj{\" o}rnsson}, {Pedersen}, {Burud},
  {Henden}, {Tanvir}, {Davis}, {Vreeswijk}, {Rol}, {Hurley}, {Cline},
  {Trombka}, {McClanahan}, {Starr}, {Goldsten}, {Castro-Tirado}, {Greiner},
  {Bailer-Jones}, {K{\" u}mmel}, \& {Mundt}}]{2001A&A...370..909J}
{Jensen}, B.~L., {Fynbo}, J.~P.~U., {Gorosabel}, J., {et~al.} 2001, \aap, 370,
  909

\bibitem[{{Kennicutt}(1998)}]{1998ARA&A..36..189K}
{Kennicutt}, R.~C. 1998, \araa, 36, 189

\bibitem[{{Koornneef}(1982)}]{1982A&A...107..247K}
{Koornneef}, J. 1982, \aap, 107, 247

\bibitem[{{Kulkarni} {et~al.}(1998){Kulkarni}, {Djorgoski}, {Ramaprakash},
  {Goodrich}, {Bloom}, {Adelberger}, {Kundic}, {Lubin}, {Frail}, {Frontera},
  {Feroci}, {Nicastro}, {Barth}, {Davis}, {Filippenko}, \&
  {Newman}}]{1998Natur.393...35K}
{Kulkarni}, S.~R., {Djorgoski}, S.~G., {Ramaprakash}, A.~N., {et~al.} 1998,
  \nat, 393, 35

\bibitem[{{Kulkarni} {et~al.}(1999){Kulkarni}, {Djorgovski}, {Odewahn},
  {Bloom}, {Gal}, {Koresko}, {Harrison}, {Lubin}, {Armus}, {Sari},
  {Illingworth}, {Kelson}, {Magee}, {Van Dokkum}, {Frail}, {Mulchaey},
  {Malkan}, {McClean}, {Teplitz}, {Koerner}, {Kirkpatrick}, {Kobayashi},
  {Yadigaroglu}, {Halpern}, {Piran}, {Goodrich}, {Chaffee}, {Feroci}, \&
  {Costa}}]{1999Natur.398..389K}
{Kulkarni}, S.~R., {Djorgovski}, S.~G., {Odewahn}, S.~C., {et~al.} 1999, \nat,
  398, 389

\bibitem[{{Lanzetta} {et~al.}(1991){Lanzetta}, {McMahon}, {Wolfe}, {Turnshek},
  {Hazard}, \& {Lu}}]{1991ApJS...77....1L}
{Lanzetta}, K.~M., {McMahon}, R.~G., {Wolfe}, A.~M., {et~al.} 1991, \apjs, 77,
  1

\bibitem[{{Lazzati} {et~al.}(2002{\natexlab{a}}){Lazzati}, {Covino}, \&
  {Ghisellini}}]{2002MNRAS.330..583L}
{Lazzati}, D., {Covino}, S., \& {Ghisellini}, G. 2002{\natexlab{a}}, \mnras,
  330, 583

\bibitem[{{Lazzati} {et~al.}(2002{\natexlab{b}}){Lazzati}, {Rossi}, {Covino},
  {Ghisellini}, \& {Malesani}}]{2002A&A...396L...5L}
{Lazzati}, D., {Rossi}, E., {Covino}, S., {Ghisellini}, G., \& {Malesani}, D.
  2002{\natexlab{b}}, \aap, 396, L5

\bibitem[{{Ledoux} {et~al.}(1998){Ledoux}, {Petitjean}, {Bergeron}, {Wampler},
  \& {Srianand}}]{1998A&A...337...51L}
{Ledoux}, C., {Petitjean}, P., {Bergeron}, J., {Wampler}, E.~J., \& {Srianand},
  R. 1998, \aap, 337, 51

\bibitem[{{Ledoux} {et~al.}(2003){Ledoux}, {Petitjean}, \& {Srianand}}]{ledoux}
{Ledoux}, C., {Petitjean}, P., \& {Srianand}, R. 2003, \mnras, 346, 209

\bibitem[{{Li} \& {Chevalier}(2001)}]{2001ApJ...551..940L}
{Li}, Z. \& {Chevalier}, R.~A. 2001, \apj, 551, 940

\bibitem[{{Lopez} \& {Ellison}(2003)}]{lopezellison}
{Lopez}, S. \& {Ellison}, S.~L. 2003, \aap, 403, 573

\bibitem[{{MacFadyen} \& {Woosley}(1999)}]{1999ApJ...524..262M}
{MacFadyen}, A.~I. \& {Woosley}, S.~E. 1999, \apj, 524, 262

\bibitem[{{Masetti} {et~al.}(2000){Masetti}, {Bartolini}, {Bernabei},
  {Guarnieri}, {Palazzi}, {Pian}, {Piccioni}, {Castro-Tirado}, {Castro Cer{\'
  o}n}, {Verdes-Montenegro}, {Sagar}, {Mohan}, {Pandey}, {Pandey}, {Bock},
  {Greiner}, {Benetti}, {Wijers}, {Beskin}, \&
  {Gorosabel}}]{2000A&A...359L..23M}
{Masetti}, N., {Bartolini}, C., {Bernabei}, S., {et~al.} 2000, \aap, 359, L23

\bibitem[{{M{\o}ller} {et~al.}(2002{\natexlab{a}}){M{\o}ller}, {Fynbo},
  {Hjorth}, {Thomsen}, {Egholm}, {Andersen}, {Gorosabel}, {Holland},
  {Jakobsson}, {Jensen}, {Pedersen}, {Pedersen}, \&
  {Weidinger}}]{2002A&A...396L..21M}
{M{\o}ller}, P., {Fynbo}, J.~P.~U., {Hjorth}, J., {et~al.} 2002{\natexlab{a}},
  \aap, 396, L21

\bibitem[{{M{\o}ller} \& {Warren}(1993)}]{1993A&A...270...43M}
{M{\o}ller}, P. \& {Warren}, S.~J. 1993, \aap, 270, 43

\bibitem[{{M{\o}ller} {et~al.}(2002{\natexlab{b}}){M{\o}ller}, {Warren},
  {Fall}, {Fynbo}, \& {Jakobsen}}]{2002ApJ...574...51M}
{M{\o}ller}, P., {Warren}, S.~J., {Fall}, S.~M., {Fynbo}, J.~P.~U., \&
  {Jakobsen}, P. 2002{\natexlab{b}}, \apj, 574, 51

\bibitem[{{M{\o}ller} {et~al.}(1998){M{\o}ller}, {Warren}, \&
  {Fynbo}}]{1998A&A...330...19M}
{M{\o}ller}, P., {Warren}, S.~J., \& {Fynbo}, J.~P.~U. 1998, \aap, 330, 19

\bibitem[{{Morton} \& {Dinerstein}(1976)}]{1976ApJ...204....1M}
{Morton}, D.~C. \& {Dinerstein}, H.~L. 1976, \apj, 204, 1

\bibitem[{{Oke} \& {Korycansky}(1982)}]{1982ApJ...255...11O}
{Oke}, J.~B. \& {Korycansky}, D.~G. 1982, \apj, 255, 11

\bibitem[{{Ostriker} \& {Heisler}(1984)}]{1984ApJ...278....1O}
{Ostriker}, J.~P. \& {Heisler}, J. 1984, \apj, 278, 1

\bibitem[{{Pei}(1992)}]{1992ApJ...395..130P}
{Pei}, Y.~C. 1992, \apj, 395, 130

\bibitem[{{Perna} {et~al.}(2003){Perna}, {Lazzati}, \&
  {Fiore}}]{2003ApJ...585..775P}
{Perna}, R., {Lazzati}, D., \& {Fiore}, F. 2003, \apj, 585, 775

\bibitem[{{Perna} \& {Loeb}(1998)}]{1998ApJ...503L.135P}
{Perna}, R. \& {Loeb}, A. 1998, \apjl, 503, L135

\bibitem[{{Petitjean} {et~al.}(2000){Petitjean}, {Srianand}, \&
  {Ledoux}}]{2000A&A...364L..26P}
{Petitjean}, P., {Srianand}, R., \& {Ledoux}, C. 2000, \aap, 364, L26

\bibitem[{{Pettini} {et~al.}(1995){Pettini}, {Hunstead}, {King}, \&
  {Smith}}]{1995qal..conf...55P}
{Pettini}, M., {Hunstead}, R.~W., {King}, D.~L., \& {Smith}, L.~J. 1995, in QSO
  Absorption Lines, Proceedings of the ESO Workshop Held at Garching, Germany,
  21 - 24 November 1994, edited by Georges Meylan. Springer-Verlag Berlin
  Heidelberg New York. Also ESO Astrophysics Symposia, 1995., p.55

\bibitem[{{Press} {et~al.}(1992){Press}, {Teukolsky}, {Vetterling}, \&
  {Flannery}}]{numrec}
{Press}, W.~H., {Teukolsky}, S.~A., {Vetterling}, W.~T., \& {Flannery}, B.~P.
  1992, {Numerical recipes in FORTRAN. The art of scientific computing}
  (Cambridge: University Press, |c1992, 2nd ed.)

\bibitem[{{Price} {et~al.}(2003){Price}, {Fox}, {Kulkarni}, {Peterson},
  {Schmidt}, {Soderberg}, {Yost}, {Berger}, {Djorgovski}, {Frail}, {Harrison},
  {Sari}, {Blain}, \& {Chapman}}]{2003Natur.423..844P}
{Price}, P.~A., {Fox}, D.~W., {Kulkarni}, S.~R., {et~al.} 2003, \nat, 423, 844

\bibitem[{{Prochaska} {et~al.}(2003{\natexlab{a}}){Prochaska}, {Gawiser},
  {Wolfe}, {Castro}, \& {Djorgovski}}]{prochaska}
{Prochaska}, J.~X., {Gawiser}, E., {Wolfe}, A.~M., {Castro}, S., \&
  {Djorgovski}, S.~G. 2003{\natexlab{a}}, \apjl, 595, L9

\bibitem[{{Prochaska} {et~al.}(2003{\natexlab{b}}){Prochaska}, {Gawiser},
  {Wolfe}, {Cooke}, \& {Gelino}}]{2003ApJS..147..227P}
{Prochaska}, J.~X., {Gawiser}, E., {Wolfe}, A.~M., {Cooke}, J., \& {Gelino}, D.
  2003{\natexlab{b}}, \apjs, 147, 227

\bibitem[{{Prochaska} \& {Wolfe}(1997)}]{1997ApJ...487...73P}
{Prochaska}, J.~X. \& {Wolfe}, A.~M. 1997, \apj, 487, 73

\bibitem[{{Prochaska} \& {Wolfe}(2002)}]{2002ApJ...566...68P}
{Prochaska}, J.~X. \& {Wolfe}, A.~M. 2002, \apj, 566, 68

\bibitem[{{Ramaprakash} {et~al.}(1998){Ramaprakash}, {Kulkarni}, {Frail},
  {Koresko}, {Kuchner}, {Goodrich}, {Neugebauer}, {Murphy}, {Eikenberry},
  {Bloom}, {Djorgovski}, {Waxman}, {Frontera}, {Feroci}, \&
  {Nicastro}}]{1998Natur.393...43R}
{Ramaprakash}, A.~N., {Kulkarni}, S.~R., {Frail}, D.~A., {et~al.} 1998, \nat,
  393, 43

\bibitem[{{Ramirez-Ruiz} {et~al.}(2002){Ramirez-Ruiz}, {Trentham}, \&
  {Blain}}]{2002MNRAS.329..465R}
{Ramirez-Ruiz}, E., {Trentham}, N., \& {Blain}, A.~W. 2002, \mnras, 329, 465

\bibitem[{{Reichart} \& {Price}(2002)}]{2002ApJ...565..174R}
{Reichart}, D.~E. \& {Price}, P.~A. 2002, \apj, 565, 174

\bibitem[{{Sari} {et~al.}(1999){Sari}, {Piran}, \&
  {Halpern}}]{1999ApJ...519L..17S}
{Sari}, R., {Piran}, T., \& {Halpern}, J.~P. 1999, \apjl, 519, L17

\bibitem[{{Sari} {et~al.}(1998){Sari}, {Piran}, \&
  {Narayan}}]{1998ApJ...497L..17S}
{Sari}, R., {Piran}, T., \& {Narayan}, R. 1998, \apjl, 497, L17

\bibitem[{{Savaglio} {et~al.}(2003){Savaglio}, {Fall}, \&
  {Fiore}}]{2003ApJ...585..638S}
{Savaglio}, S., {Fall}, S.~M., \& {Fiore}, F. 2003, \apj, 585, 638

\bibitem[{{Schaye}(2001)}]{2001ApJ...562L..95S}
{Schaye}, J. 2001, \apjl, 562, L95

\bibitem[{{Schlegel} {et~al.}(1998){Schlegel}, {Finkbeiner}, \&
  {Davis}}]{1998ApJ...500..525S}
{Schlegel}, D.~J., {Finkbeiner}, D.~P., \& {Davis}, M. 1998, \apj, 500, 525

\bibitem[{{Shapley} {et~al.}(2003){Shapley}, {Steidel}, {Pettini}, \&
  {Adelberger}}]{2003ApJ...588...65S}
{Shapley}, A.~E., {Steidel}, C.~C., {Pettini}, M., \& {Adelberger}, K.~L. 2003,
  \apj, 588, 65

\bibitem[{{Silva} \& {Viegas}(2002)}]{2002MNRAS.329..135S}
{Silva}, A.~I. \& {Viegas}, S.~M. 2002, \mnras, 329, 135

\bibitem[{{Solomon} {et~al.}(1987){Solomon}, {Rivolo}, {Barrett}, \&
  {Yahil}}]{1987ApJ...319..730S}
{Solomon}, P.~M., {Rivolo}, A.~R., {Barrett}, J., \& {Yahil}, A. 1987, \apj,
  319, 730

\bibitem[{{Stanek} {et~al.}(2003){Stanek}, {Matheson}, {Garnavich}, {Martini},
  {Berlind}, {Caldwell}, {Challis}, {Brown}, {Schild}, {Krisciunas}, {Calkins},
  {Lee}, {Hathi}, {Jansen}, {Windhorst}, {Echevarria}, {Eisenstein}, {Pindor},
  {Olszewski}, {Harding}, {Holland}, \& {Bersier}}]{stanek}
{Stanek}, K.~Z., {Matheson}, T., {Garnavich}, P.~M., {et~al.} 2003, \apjl, 591,
  L17

\bibitem[{{Steidel} \& {Sargent}(1992)}]{1992ApJS...80....1S}
{Steidel}, C.~C. \& {Sargent}, W.~L.~W. 1992, \apjs, 80, 1

\bibitem[{{Tumlinson} {et~al.}(2002){Tumlinson}, {Shull}, {Rachford},
  {Browning}, {Snow}, {Fullerton}, {Jenkins}, {Savage}, {Crowther}, {Moos},
  {Sembach}, {Sonneborn}, \& {York}}]{2002ApJ...566..857T}
{Tumlinson}, J., {Shull}, J.~M., {Rachford}, B.~L., {et~al.} 2002, \apj, 566,
  857

\bibitem[{{Turnshek} {et~al.}(1989){Turnshek}, {Wolfe}, {Lanzetta}, {Briggs},
  {Cohen}, {Foltz}, {Smith}, \& {Wilkes}}]{1989ApJ...344..567T}
{Turnshek}, D.~A., {Wolfe}, A.~M., {Lanzetta}, K.~M., {et~al.} 1989, \apj, 344,
  567

\bibitem[{{Van Dokkum}(2001)}]{2001PASP..113.1420V}
{Van Dokkum}, P.~G. 2001, \pasp, 113, 1420

\bibitem[{{Viegas}(1995)}]{1995MNRAS.276..268V}
{Viegas}, S.~M. 1995, \mnras, 276, 268

\bibitem[{{Vladilo} {et~al.}(2001){Vladilo}, {Centuri{\' o}n}, {Bonifacio}, \&
  {Howk}}]{2001ApJ...557.1007V}
{Vladilo}, G., {Centuri{\' o}n}, M., {Bonifacio}, P., \& {Howk}, J.~C. 2001,
  \apj, 557, 1007

\bibitem[{{Vreeswijk} {et~al.}(2003){Vreeswijk}, {Wijers}, {Rol}, \&
  {Hjorth}}]{2003GCN..1953....1V}
{Vreeswijk}, P., {Wijers}, R., {Rol}, E., \& {Hjorth}, J. 2003, GRB Circular
  Network, 1953

\bibitem[{{Vreeswijk} {et~al.}(2001){Vreeswijk}, {Fruchter}, {Kaper}, {Rol},
  {Galama}, {Van Paradijs}, {Kouveliotou}, {Wijers}, {Pian}, {Palazzi},
  {Masetti}, {Frontera}, {Savaglio}, {Reinsch}, {Hessman}, {Beuermann},
  {Nicklas}, \& {Van den Heuvel}}]{2001ApJ...546..672V}
{Vreeswijk}, P.~M., {Fruchter}, A., {Kaper}, L., {et~al.} 2001, \apj, 546, 672

\bibitem[{{Wolfe}(1987)}]{1987txra.symp..309W}
{Wolfe}, A.~M. 1987, in 13th Texas Symposium on Relativistic Astrophysics,
  309--313

\bibitem[{{Wolfe} {et~al.}(1995){Wolfe}, {Lanzetta}, {Foltz}, \&
  {Chaffee}}]{1995ApJ...454..698W}
{Wolfe}, A.~M., {Lanzetta}, K.~M., {Foltz}, C.~B., \& {Chaffee}, F.~H. 1995,
  \apj, 454, 698

\bibitem[{{Wolfe} {et~al.}(1986){Wolfe}, {Turnshek}, {Smith}, \&
  {Cohen}}]{1986ApJS...61..249W}
{Wolfe}, A.~M., {Turnshek}, D.~A., {Smith}, H.~E., \& {Cohen}, R.~D. 1986,
  \apjs, 61, 249

\bibitem[{{Woosley}(1993)}]{1993ApJ...405..273W}
{Woosley}, S.~E. 1993, \apj, 405, 273

\end{thebibliography}

\end{document}